\documentclass{aa}  

\usepackage{graphicx}
\usepackage{txfonts}
\usepackage[colorlinks=true,citecolor=blue]{hyperref}
\usepackage{breqn}

\usepackage{silence}
\newcommand{\Msunyr}{M_\odot \, \textup{yr}^{-1}}
\newcommand{\kms}{km \, \textup{s}^{-1}}

\DeclareRobustCommand{\rchi}{{\mathpalette\irchi\relax}}
\newcommand{\irchi}[2]{\raisebox{\depth}{$#1\chi$}}
\begin{document}

   \title{ The role of heating in the formation and the dynamics of YSO jets}

   \subtitle{I. A parametric study}

   \author{C. Meskini
          \inst{1}
          \and
          C. Sauty
          \inst{1}\fnmsep\inst{2}
          \and
          A. Marcowith
          \inst{1}
            \and
          N. Vlahakis
          \inst{3}
          \and
          V. Brunn
          \inst{1}}
   \institute{Laboratoire Univers et Particules de Montpellier, Université de Montpellier/CNRS, place E. Bataillon, cc072, 34095 Montpellier, France \\
   \email{chadi.meskini@umontpellier.fr}
         \and
             Laboratoire Univers et Th\'eories, Observatoire de Paris, Université PSL, Universit\'e Paris Cit\'e, CNRS, F-92190 Meudon, France
         \and
             Department of Physics, National and Kapodistrian University of Athens, University Campus, Zografos GR-157 84 Athens, Greece
             }

   \date{Received 12 January 2024 / accepted 01 April 2024}

% \abstract{}{}{}{}{} 
% 5 {} token are mandatory
 
  \abstract
   {Theoretical arguments as well as observations of young stellar objects (YSOs) support the presence of a diversified circumstellar environment. A stellar jet is thought to account for most of the stellar spin down and disk wind outflow for the observed high mass-loss rate, thus playing a major role in the launching of powerful jets. RY Tau, for instance, is an extensively studied intermediate mass pre-main sequence star. Observational data reveal a small-scale jet called micro-jet. Nevertheless, it is not clear how the micro-jet shapes the jet observed at a large scale.}
   {The goal is to investigate the spatial stability and structure of the central jet at a large scale by mixing the stellar and disk components.}
   {two existing analytical self-similar models for the disk and the stellar winds to build the initial setups. Instead of using a polytropic equation of state, we mapped the heating and cooling
sources from the analytical solutions. The heating exchange rate was controlled by two parameters, its spatial extent and its intensity.}
   {The central jet and the surrounding disk are strongly affected by these two parameters. We separate the results into three categories, which show different emissivity, temperature, and velocity maps. We reached this categorization by looking at the opening angle of the stellar solution. For cylindrically, well-collimated jets, we have opening angles as low as $10^{\circ}$ between $8-10 \ au$, and for the wider jets, we can reach $30^{\circ}$ with a morphology closer to radial solar winds.}
   {Our parametric study shows that the less heated the outflow is, the more collimated it appears. We also show that recollimation shocks appear consistently with UV observations in terms of temperature but not density.}


   \maketitle
%
%________________________________________________________________

\section{Introduction}
Pre-main sequence stars, particularly Class II classical T Tauri stars (CTTSs), are in an active evolutionary stage with ongoing accretion and ejection processes \citep{ray2007protostars,bally2007observations,garufi2019sphere,sauty2019jet,ray2021jets}. A substantial fraction of CTTSs present observational signatures of outflows, and more than 60\% exhibit stellar winds \citep{Kwanetal07}.
CTTSs show a wide interval of rotation rates. However, the trend for the rotation speed of these stars stay way below the break-up velocity \citep{bouvier1986rotation,bertout1989t,bouvieretal93,rebull2004stellar,herbst2007protostars}. Although CTTSs are actively accreting mass from the disk with mass accretion rates around  $\sim 10^{-10}-10^{-8} \ \Msunyr$, observations indicate that the rotational velocities of CTTSs stay constant over a few million years. 

The resulting jets are supersonic, collimated outflows that are closely linked to accretion. Observations establish a correlation between accretion and ejection, which clearly shows that the disk is an essential ingredient in jet formation \citep{cabrit1990forbidden,hartiganetal95,mirabel1998accretion,Manara2016Evidence}. It was found that at early stages of a young star life, the strength of the winds and the jets are more powerful. Such an observation is explained by the vigorous matter infall phase early in the evolutionary stage of the star \citep{andre1992vlbi}.

The presence of a jet is determined by the existence of a surrounding accretion disk. No disk or no disk-star interaction implies no observable jet. This is confirmed by the absence of detectable jets in weak-lined T Tauri stars (WTTS). Nonetheless, the opposite is not necessarily true \citep{ray2021jets}. A jet may originate from different sources -- the star, the circumstellar accretion disk, or from the star-disk interaction (hereafter SDI) region where the stellar magnetosphere connects to the disk. Some jets appear in the form of sporadic non-collimated ejections. Overall, arguments on the jet nature support the idea that there must be more than one mechanism to explain the outflows. Although, the outflows can extend from a few hundred au to the parsec scale, the main region of jet collimation and acceleration is below $\sim 100$ au \citep{bally2007observations,petrovetal19,takami2022time}.

To explain the observed low rotational velocities,  angular momentum must be removed from the star and its surrounding accretion disk. Many models have been developed to explain the mechanism responsible for angular momentum transport. Hydrodynamical models alone fail to explain young
stellar object (YSO) jets. Adding a magnetic field is a promising mechanism that matches the observed jet collimation and ejection efficiency \citep{cabrit2007jets}.

As for the solar wind \citep{pantellini1988velocity}, pure thermal coronal wind models fail to explain jets with high enough mass-loss rates because they would require higher temperatures than is observed \citep{decampli1981t}. 
In fact, if we consider the high temperatures coupled with the high densities necessary for jets, the resulting X-ray emissions would be in excess with respect to the data. \cite{bisnovatyi1977models} estimated X-ray luminosities on the order of $10^{34} \ \rm{erg  \ s^{-1}}$ for a spherical model of thermally driven coronal wind. On the other hand, observed luminosities emanating from T Tauri stars are $L_X \leq 10^{30} \ \rm{erg \ s^{-1}}$ \citep{feigelson1999high,imanishi2003systematic}. Invoking Alfv\'en waves, so as to support coronal heating, would alleviate the X-ray luminosity excess. Alone, Alfv\'en waves would require a strong luminosity. Although good candidates for energy transport, the non-dissipative nature of these waves makes it difficult to extract the embedded mechanical energy in the form of heat. X-ray emission plus Alfv\'en waves can support ejections close to $10^{-9} \Msunyr$, and for temperatures around one million degrees the jets can reach velocities up to 300 $\kms$ \citep{decampli1981t}. Another source of heating could be turbulent ram or magnetic pressure. Thus turbulent pressure \citep{hartmann1982wave} or accretion-powered stellar wind (APSW)  scenarios \citep{matt2005accretion} have the potential to produce sufficient mass-loss rates. For some parameters in accretion-powered wind, the model also results in magnetic braking  spinning the star down. The decrease in rotational velocity occurs in the lifetime of the star \citep{matt2012magnetic}.

Nonetheless, thermal winds can be invoked to launch micro-jets with low mass-loss rates of a few $10^{-9} \Msunyr$, such as the RY Tau micro-jet observed by \cite{StongeBastien08} with $H\alpha$ tracers \citep{sautyetal11, sauty2022nonradial}. The stellar wind is strong enough to brake the star in its lifetime. 

Magnetocetrifugal driving, for instance, has the advantage of keeping the outflows corotating with the star using magnetic torque to carry off a substantial fraction of the angular momentum. Yet this model lacks efficient ways of launching plasma out of the star.

The most efficient mechanism to explain collimated outflows is the presence of large-scale magnetic fields combined with a rotating disk \citep{casse2000magnetized}. Large-scale magnetic fields can be created during the contraction of the primordial envelope, thus advecting the magnetic field, or by the dynamo effect \citep{mouschovias1976nonhomologous,rekowski2000structure}. Since the work by \cite{blandford1982hydromagnetic} (hereafter BP82), it has been known that a disk threaded by a magnetic field can accelerate material if the latter crosses all critical surfaces. The model presented in BP82, based on the work of \cite{bardeen1978model}, uses a self-similar approximation to solve the magnetohydrodynamical (MHD) equations. Nevertheless, self-similarity could impact the jet dynamics given that the axis of rotation as well as the interaction with the ambient medium are not taken into account. However, self-similarity is a good proxy for objects where the rotational axis and the jet are almost parallel. It also shortens the list of free parameters, and thus helps for the exploration of more space parameters in regards to the magnetic field distribution (e.g., \citealt{contopoulos1994magnetically, ferreira1997magnetically}) and the thermal effects (e.g., \citealt{vlahakis2000disc, casse2004radiatively, jannaud2023numerical}). 

A numerical study of a jet launching and collimation can be tackled in two main ways: either one can consider the disk as a boundary condition  (\citealt{ustyugova1995magnetohydrodynamic,ustyugova1999magnetocentrifugally,koldoba1995simulations,ouyed1997numericala,pudritz2006controlling,porth2010acceleration,jannaud2023numerical}; and references therein) or one can include the complete physics of the disk to investigate the feedback of the disk structure on the jet launching (e.g., \citealt{casse2002magnetized,casse2004radiatively,zanni2007mhd,tzeferacos2009magnetization,tzeferacos2013effects,fendt2013bipolar,jacquemin2021magnetic,takasao2022three}). These simulations used a turbulent diffusivity due to the development and saturation of magnetorotational instability in the disk. Thus the authors used an $\alpha$-disk prescription for the viscosity and magnetic diffusivity following Shakura-Sunyaev. \cite{mattia2020magnetohydrodynamica,mattia2020magnetohydrodynamicb, mattia2022jets} have explored more realistic diffusion coefficients including a full study of the dynamo development and saturation in the disk. In \cite{mattia2022jets}, the authors show how the diffusive quenching of the dynamo, taking the feedback of the large dynamo magnetic field, stabilizes the system. A stronger quenching of the dynamo leads to lower magnetization.

Each of these methods has its drawbacks. While platform simulations (using the disk as a boundary) enable the study of jets at very large scales, they have extensive degrees of freedom. Several distributions have to be specified at the wind injection boundary. Consequently, inferring a set of generic results on collimation has proven difficult.
On the other hand, including the disk substantially increases the computation time. To mitigate a long computation time, we had to use a lower resolution. In turn, smaller domains of parameters were studied.

Before including the disk physics in a future work, for the present publication, we used platform simulations to investigate jet launching and collimation by choosing computational domains bigger than the ones used in \cite{ZanniFerreira13}, \cite{Irelandetal2021}, and \cite{ireland2022effect}, but smaller than the ones in \cite{Matsakosetal2009}, \cite{stute20143d}, and \cite{jannaud2023numerical}.
We used semi-analytical models for the disk \citep{vlahakis2000disc} (ADO), and the model from \cite{sautyetal11} for the stellar component (ASO). We followed the procedure of \cite{sauty2022nonradial} to include a dead zone, which is a static stellar atmosphere, and an accretion zone. Both structures are not present in the original stellar model and have been tested in regards to the stability in \cite{sauty2022nonradial}.  

The presence of a hot corona above the disk surface heats the base of the disk wind, significantly increasing the outflow mass \citep{casse2000magnetized,bai2016magneto}.
\cite{casse2000magnetized} showed that adding a similar heating source to what is emitted by an embedded source (i.e., a star) counterbalances the magnetic compression, enabling smaller mass loading. This illustrates the importance of heating in jet formation. We follow here this framework by including a heating rate in the stellar and disk wind components to infer the jet properties, that is, the degree of collimation, velocities, mass-loss rates, temperatures, etc.

This work is the first of a series aiming to explore the effects of extra sources of heating, besides magnetocentrifugal driving, for jet launching and the production of stable collimated jets. Our approach uses a parameterized heating to quantitatively determine the amount of heating rate needed to launch and maintain a jet.

The paper is organized as follows. Sec.\ref{sec:num setup} describes the ideal MHD equations and the analytical solutions used at initialization. The stellar analytical solution surrounded by the disk wind solution were used to map the initial conditions. In Sec.\ref{sec: 3}, the normalization factors as well as the mixing and heating functions are introduced. We show, in Sec.\ref{sec:results}, the effect of heating on the jet dynamics, that is, the degree of collimation, speed, velocities, etc. Namely, as we heated up the disk atmosphere, the jet became wider. On the other hand, if we decreased the heating available, we got a propeller regime \citep{romanovaetal09, romanovaetal11}, where magnetospheric ejections affect the jet. In Sec.\ref{sec:discussion}, we discuss which configurations are closest to the observations. We then explore possible heating processes. 

\section{MHD equations and analytical solutions} \label{sec:num setup}
\subsection{Physical framework and MHD equations}
We intend to study properties and stability of magnetically driven disk outflows which will be referred to as " disk jet" hereafter. We call "spine" jet the plasma ejected from the star, as well as the plasma from the magnetosphere heated by the stellar corona up to the SDI zone. We use the same nomenclature that was used in \cite{jannaud2023numerical} to clearly differentiate between the contribution of each component. 
To keep track of the separation, we use a passive tracer $C_{Tr}$ that canvas the evolution of each component, for which an example is given in Appendix \ref{tarcers},
\begin{equation}
    \frac{\partial (\rho C_{Tr})}{\partial t} + \nabla \cdot (\rho C_{Tr} {\bf V}) = 0.
\end{equation}

We use MHD equations, following the evolution of the heaviest species which dominate in the plasma. The plasma of interest is partially ionized, but still can be described by a one fluid approximation. Ionization has to be high enough for the magnetic field to play a role in the plasma evolution. 

The disk itself is set as a boundary condition with a Keplerian profile.
Both the central object and the disk are assumed to be threaded by a large scale magnetic field. This is not the case of \citet{zanni2009mhd,romanovaetal09,ZanniFerreira13}, where they use a stellar dipolar field that thread the accretion disk. In order to describe such a flow we use the ideal MHD equations that express the conservation of macroscopic fluid quantities,
\begin{equation}\label{mass conservation}
    \frac{\partial \rho}{\partial t} + \nabla \cdot (\rho {\bf V}) = 0,
\end{equation}
\begin{equation}\label{momentum conservation}
    \rho \left[\frac{\partial {\bf V}}{\partial t} + ({\bf V} \cdot \nabla){\bf V}\right] = -\nabla P+ \frac{c}{4 \pi} (\nabla \times {\bf B}) \times {\bf B} - \rho \frac{\mathcal{G} \mathcal{M}}{r^2}{\bf \hat r},
\end{equation}
\begin{equation}\label{magnetic induction}
    \frac{\partial {\bf B}}{\partial t} - \nabla \times ({\bf V} \times {\bf B}) = 0,
\end{equation}
\begin{equation}\label{energy equation}
    \frac{\partial P}{\partial t} + \mathbf{V} \cdot \nabla P + \Gamma P \nabla \cdot \mathbf{V} = Q ,
\end{equation}
where $\rho$, $P$, and $\mathbf{V}$ are the plasma density, pressure and velocity, respectively. Finally, $\bf B$ refers to the magnetic field. Hereafter, for clarity, we have adopted the following notations. Subscripts $d$ and $s$ are for the Disk and Stellar solutions, respectively, and the $\star$ indicates quantities at the Alfv\'en location along a given fieldline for a given model. We also note that ($r$, $\theta$, $\phi$) and ($\varpi$, $\phi$, $z$) are spherical and cylindrical coordinates, respectively. $\theta$ in spherical coordinates is the colatitude. We solve the MHD equations using spherical coordinates. A detailed description of the computational grid is found in Sec.\ref{sec:grid and initial conditions}.

Since axisymmetry is assumed, mass conservation (Eq.\ref{mass conservation}) is solved using the poloidal components of the velocity, $V_r$ and $V_\theta$, respectively. The left hand side of Eq.\ref{momentum conservation} describes acceleration, while the right hand side is the sum of hydrostatic effects, Lorentz force expressed as $(\nabla \times {\bf B}) \times {\bf B}$, and the gravitational force, with $\mathcal{G}$ the gravitational constant, and $\mathcal{M}$ the stellar mass. The induction equation (Eq.\ref{magnetic induction}) describes the temporal evolution of the magnetic field where magnetic resistivity is neglected since we assume ideal MHD.
Finally, Eq.\ref{energy equation} describes the time dependent energy equation. $Q$ is the volumetric heating rate accounting for energy gains and losses. We use $Q = (\Gamma - 1)q$, with $q$ the volumetric energy source term, and $\Gamma$ the adiabatic index, which has the value of $5/3$ for a monoatomic gas.

Different source terms are used for each component, keeping in mind the initial pressures and velocities that we use are computed from the analytical solutions. The disk wind can be described by an adiabatic or isothermal  evolution through the polytropic index $\gamma$. To implement such a behavior we will use the source term presented in \cite{Matsakosetal2008,Matsakosetal2009},
\begin{equation}\label{ADo heating}
    Q = (\Gamma - \gamma)P(\nabla \cdot \mathbf{V}),
\end{equation}
with $\gamma$ the effective polytropic index. In the case of adiabatic evolution, $\gamma=\Gamma=5/3$. The entropy, which is proportional to $P/\rho^\Gamma$ is conserved along each streamline. This equation ensures that the pressure does not take negative values. When the system is not adiabatic, $\gamma \neq \Gamma$ and the heating rate is not zero. We calculate it with Eq. \ref{ADo heating} in the disk wind solution, using $\gamma = 1.05$. In that case, Eq.\ref{ADo heating} is equivalent to writing $P \propto \rho^\gamma$.

As for the stellar jet, the analytical solution describes dense jet with an under-pressured structure as in \cite{sautyetal2002}. The stellar jet has a lower density along the axis than at its edges. Conversely, in the works mentioned above, the stellar contribution is either very weak compared to the disk wind or is completely neglected. The main physical variables have been constrained from observations \citep{sautyetal11}. For the jet to maintain physical mass loss rates we have to use a higher energy source term than the polytropic equation offers. One could play on polytropic index $\gamma$ to toggle heat exchange in the stellar component. If not enough heating is present in the stellar component then simulations show a collapse of the jet. An adiabatic evolution results in the jet turning into a static atmosphere, while isothermal conditions produce a weak turbulent jet (e.g., \citealt{Matsakosetal2009,stute20143d}). The stellar analytical solution is computed given a heat rate along the flow. This rate is given by,
\begin{equation}\label{ASO heating}
    Q = \rho {\bf V} \cdot \left[\Gamma \nabla \left(\frac{P}{\rho}\right) - (\Gamma-1) \frac{\nabla P}{\rho}\right].
\end{equation}

First, Eq.\ref{ASO heating} is equivalent to Eq.\ref{energy equation} in the steady case.
Eq.\ref{ASO heating} characterizes the heat rate deposit in a comoving frame with the fluid. The energy equation has to be proportional to the velocity. However, changes in direction and magnitude are not considered outside the accretion region. As a consequence, the spine jet solution no longer starts at an equilibrium state. Nevertheless, since we use a relaxation method from a heat map that does not vary from its initial structure, we do not account for velocity changes in the time dependent energy equation. The poloidal velocity sign is reversed in the accretion funnel, and observed accretion mass flux is reproduced by multiplying velocity and density by constant factors. Subsequently, modifying velocity has to be reflected onto the heating rate by introducing the same constant to the quantity. This keeps a self-consistent description of the accretion region. The inflow is sub-Alfv\'enic in the accretion region, which means the solution starts close to an equilibrium state. For an in depth explanation refer to Sec.2.2 of \cite{sauty2022nonradial}.

\subsection{Analytical solutions}
\subsubsection{Spine jet solution, ASO}
In order to model the RY Tau micro-jet, \cite{sautyetal11} applied a semi-analytical solution derived from the self-similar approach introduced by \cite{sautyTsinganos94}. Crafted for jets originating from low-mass accreting T Tauri stars (TTS) they constitute an exact solution to the steady ideal MHD equations. This solution describes density ($\rho$), pressure ($P$), velocity field ($V$), and magnetic field ($B$) of a mono-fluid. These equations govern the behavior of a fully ionized plasma composed of protons and electrons only around a TTS.

We use the solution presented in \cite{sautyetal11}, the stellar parameters of RY Tau for mass and radius, respectively, $M = 1.6 M_\odot$, $R = 2.4 R_\odot$ \citep{hartiganetal95}. For the mass-loss rate, we use $\dot{M} = 10^{-8.5} \Msunyr$ \citep{gomezetal01}. More recent observations propose higher masses and radii (e.g., \citealt{calvetetal04,dotter2008dartmouth,bressan2012parsec,dotter2016mesa}), with a mass centered around $2 M_\odot$, and a radius varying between $2.37$ and $3.7 R_\odot$ for RY Tau \citep{davies2020inner}. While $1.6M_\odot$ is certainly below the lower limit, a change in mass from $1.6 M_\odot$ to $2 M_\odot$ does not affect the solution. A scaling of the velocity, magnetic field, and mass flux to match higher masses would increase these quantities by $10\%$. At the stellar surface, the equatorial rotational velocity given by the solution is $8.6 \ \rm{km \ s^{-1}}$ \citep{sautyetal11}, corresponding to a period of 14 days, which is the value present in the simulations. RY Tau is a faster rotator with a period closer to 3 days, if an inclination of $66^{\circ}$ is taken into account in the projected velocity. The difference in rotational velocity between observations and our simulations can be overlooked for now as the stellar outflow is mainly pressure driven which lessens the importance of the rotational output of the analytical solution on the dynamics of the jet.

Overall, the stellar component is within  the constraints set by \cite{gomezetal01,gomezetal07}, \cite{StongeBastien08}, and \cite{agra-amboageetal09}. The specific stellar solution we used is tailored for the large scales and small scales of RY Tau jet. To study other objects we can either compute new stellar solutions with different stellar parameters or scale the solution presented above to a star with a different mass. However, \cite{albuquerque2020accretion} showed that a simple rescaling of the solution is usually not sufficient to be in agreement with all physical parameters. To make  simulations of a jet from another CTTS, we would need to produce new self-similar solutions as initial conditions.

Further modifications have been applied to the spine jet solution in order to take into account physical features not contained in the initial model. In addition to a micro-jet, a second region between $0.03$ and $0.09 \ au$ on the equatorial plane was adapted to describe an accretion region. To obtain such a region, density, and velocity were multiplied by constant factors as to inverse the flow of material and increase accretion mass flux. Physically, the accretion speed is close to $290 \ \rm{km \ s^{-1}}$, and accretion rate is in the order of $10^{-8}\Msunyr$. A static corona was added between the star and the accretion zone, where the poloidal velocity is set to zero. Thus no angular momentum is transported in this region. For more details about the consequences of varying the size and density of the accretion regions, see \cite{sauty2022nonradial}. We use the same densities and velocities in the accretion region as testC and testD from the above paper to compare the evolution of the stellar solution with a more realistic disk wind.

\subsubsection{Disk wind solution, ADO}
Disk winds originate from streamlines that reach the surface of the disk. From that point on they follow the magnetic field and cannot cross them. Faraday's induction law dictates that an electromotive force across the disk creates a differential electric potential, and by closing the electric circuit (see lecture notes from chapter on MHD disk winds in \citealt{ferreira2007jets}), 
a radial Lorentz force appears. This force induces, at the same time, a slowing down of the disk as expected from Lenz's law. To collimate the jet the azimuthal component of the magnetic field is crucial. By coiling along the jet, the magnetic field creates an inward force that collimates the jet.  
If a horizontal cut is made at some distance, $z$, and if the current is non-vanishing, then the Laplace force is directed toward the jet axis. \cite{blandford1982hydromagnetic} originally proposed the first MHD radial self-similar solutions describing a cold disk wind. 

Their work was, later, extended to fill in for some of the models short comings. \cite{ferreira1997magnetically} (referred to as F97 hereafter) consistently connect disk and disk wind. \cite{vlahakis2000disc} (referred to as VTST00 hereafter) on the other hand present a model, with a thermal component, crossing all critical surfaces causally disassociating the outflow from its origin. 
%\sout{(\cite{vlahakis2000disc}; referred to as VTST00 hereafter)}.
As we do not implement the physics of the accretion disk and use it as a boundary condition, we will rather use the VTST00 self-similar solution, because it crosses all three critical surfaces. In order to match the MHD wind solution with the resistive accretion disk in a consistent way, one has to avoid discontinuities, which put extra constraints on the wind. Such an investigation is postponed to future work.

To describe the disk wind solution we have to specify eight physical quantities ($P  , \ \rho  , \ \mathbf{V}  , \ \mathbf{B}$) at the disk surface. The self-similar ansatz to determine these quantities is presented in the Appendix B. VTST00 built a model with a finite temperature and a solution that starts at a subslow magnetosonic velocity. The solution then crosses the slow magnetosonic, Alfv\'en, and fast magnetosonic critical surfaces. The authors discuss the implication of a slightly higher magnetic field distribution on crossing critical points, and point out the minimal effect it may have on the resulting solution. The magnetic scaling parameter, $x$, which controls the magnetic field distribution, can be directly linked to the ejection index, $\xi$, as expressed in \cite{ferreira1997magnetically} ($\xi = 2(x-3/4)$). This parameter defines the disk accretion rate as $\dot{M} \propto r^\xi$ \citep{ferreira1995magnetized} and mass flux in the wind for a disk with a radially self-similar structure. We choose a value from BP82, $x=3/4$, which corresponds to $\xi = 0$. In that case $\dot M$ varies logarithmically with $\varpi$.

\cite{ferreira1997magnetically} demonstrates that for ejection to happen $\xi$ has to be higher than 0, and as $\xi$ grows larger, so does the ejection efficiency. 
%$\xi$ should remain close to zero. NO ONLY IF THERE IS ONLY ACCRETION POWER
We argue that since VTST00 show almost identical behavior for a solution with a higher ejection index, the solution should not go against the theoretical arguments put forward by \cite{ferreira1997magnetically} (see discussion below). The constraint present in F97 arises from linking disk and wind, a condition absent in the chosen disk model. 
Having a consistent disk seems to come at a price, as solutions crossing the fast surface are hard to find.

Although, we have chosen $\xi \approx 0$, this does not mean we do not have mass losses from the disk. The model adds a thermal component that helps material travel outside the disk. To further understand the effect due to this parameter, we vary its value close to $3/4$ and notice no quantitative difference. Testing with $x=0.7575$, above what was probed by VTST00, we obtain small differences on scales not yet achieved by the current instruments. A higher ejection efficiency does not change the external equilibrium. Thus, linking the accretion disk and its wind is not a fundamental issue in this paper. This being said, other solutions with different $\xi$ are going to be chosen in a future work.

We focus here on the physical implications of using the solution presented in VTST00. The ratio of the Alfv\'en poloidal speed to the Keplerian velocity at the disk boundary level results in a value of $\sim 0.014$. Such a value means that at the disk surface the energy density of the magnetic field is less than the kinetic energy from rotation, implying the magnetic field follows the plasma. 

Then, looking at the ratio of the sound speed and the initial velocity we deduce that the ejection velocity at the base of the flow is negligible compared to the sound speed. The disk wind solution of VTST00 is magnetically driven but there is a heating source above the disk. It helps the initial acceleration of the flow.  The Keplerian velocity is around two times larger than the thermal speed ($C_s = \sqrt{\partial P / \partial \rho}$). Consequently, the disk rotation cannot be suppressed by thermal effects. The terminal velocity of the solution is $\sim 400 \ \rm{km \ s^{-1}}$ which is well within the observational range of YSOs. VTST00 estimated the mass-loss rate for a field anchored at $0.6 \ \rm{au}$ with a magnetic field of $8~\rm{G}$. They estimate mass loss at $ 10^{-8} \ \Msunyr$ with a temperature of $8 \times 10^3 \ K$. Finally, the specific angular momentum carried by the wind to the angular momentum needed for accretion ratio is $\leq 1.5\%$. Since the outflowing plasma is at most $1\%$ the value of accreted materiel, this implies all the angular momentum is carried by the wind.

\subsection{Mixing analytical solutions}

In \cite{sauty2022nonradial}, the authors show that the inner analytical solution of \cite{sautyetal11} is quite stable in the inner spine jet region.  Even with the introduction of an accretion region, simulations and analytical work give similar stellar spin-downs. Both angular momentum and mass-loss rates extracted by the stellar jet match values given by the analytical work \citep{sautyetal11}. The stellar braking time was computed to be approximately 0.6 million years. The convergence of these results between the analytical solution and the numerical simulations can be attributed to two primary factors. First, as described in \cite{sautyetal11}, only the open field-line region of the stellar jet is responsible for the stellar braking. Second, reversing the sign of the toroidal magnetic field ($B_\phi = -B_{\phi,analytic}$) and the poloidal velocity ($V_p = -1.5 V_{p,analytic}$) inside the magnetospheric accretion zone guaranties the azimuthal velocity stays unchanged on the star and consistent with isorotation law. Moreover the density was multiplied by a factor $5$ to increase the accretion rate to $\sim 10^{-8}\Msunyr$ \citep{sauty2022nonradial}. See Fig.1 in this publication where we kept the initial conditions in domains 1 (the stellar jet component), 2 (the magnetospheric accretion columns), 3 (the magnetostatic dead zone), and changed them in region 4 (the disk wind domain). In essence, the disk effectively exerts magnetic braking on the star as the magnetospheric field lines efficiently lock the star to the disk. This way of treating the accretion zone is kept in our study.
Consequently, the simulations demonstrate that the stellar jet and magnetospheric accretion collectively contribute to the star spin-down in a manner resembling of \cite{sautyetal11}, resulting in a spin-down period of less than one million years for the star $-$ a timescale well within the lifetime of the CTTS phase.

We extend the work simulating RY Tau environment \citep{sauty2022nonradial} by adding a more realistic disk wind. The one used in their study is not dense enough, resulting in a low mass loss from the disk. Furthermore, we go beyond studying the jet dynamics above the fast magnetosonic surface as in \citep{Matsakosetal2009,stute2008stability}, by including the launching region. This, in turn, will enable us to build consistently the launching region in the disk and its interaction with the stellar component.

\section{From analytical solutions to numerical simulations} \label{sec: 3}
\subsection{Grid and initial conditions}\label{sec:grid and initial conditions}
To solve our system of equations we use $PLUTO$, a code based on finite volume methods \citep{mignone2007pluto}. Spatial reconstruction of primitive quantities is achieved with linear interpolation \citep{mignone2014high}. For flux computation we choose a robust solver, HLL (for Hartan Lax, Van Leer approximate Riemann). Finally, to ensure the divergence-free condition $\nabla \cdot {\bf B}=0$ we use the eight-wave structure \citep{powell1997approximate}.

The simulations presented below are 2.5D. We consider a 3D grid while assuming axisymmetry around the stellar axis of rotation. The grid is spherical with ($R$, $\theta$) as coordinate components.

The equatorial plane has a planar symmetry, and the colatitude $\theta$ is discretized into 128 points starting from $\theta \sim 0$, at the polar axis, to $\theta = 1.552$.
This limit arises when $PLUTO$ creates a ghost zone that reaches the equatorial plane. At the equator, the analytical equations used to initialize the disk wind diverge. Hence, the system is not solved beyond $\theta_{lim}=1.552$, unless we increase the number of points in colatitude to decrease the size of the ghost zones. 
But this increase in turn decreases the time step of a simulation. The right compromise between resolution and time convergence is obtained by using 128 discretization points. Adding more points does not affect the global evolution or the steady state solutions if reached.

\begin{figure}[htbp]
    \centering
    \includegraphics[width=.45\textwidth]{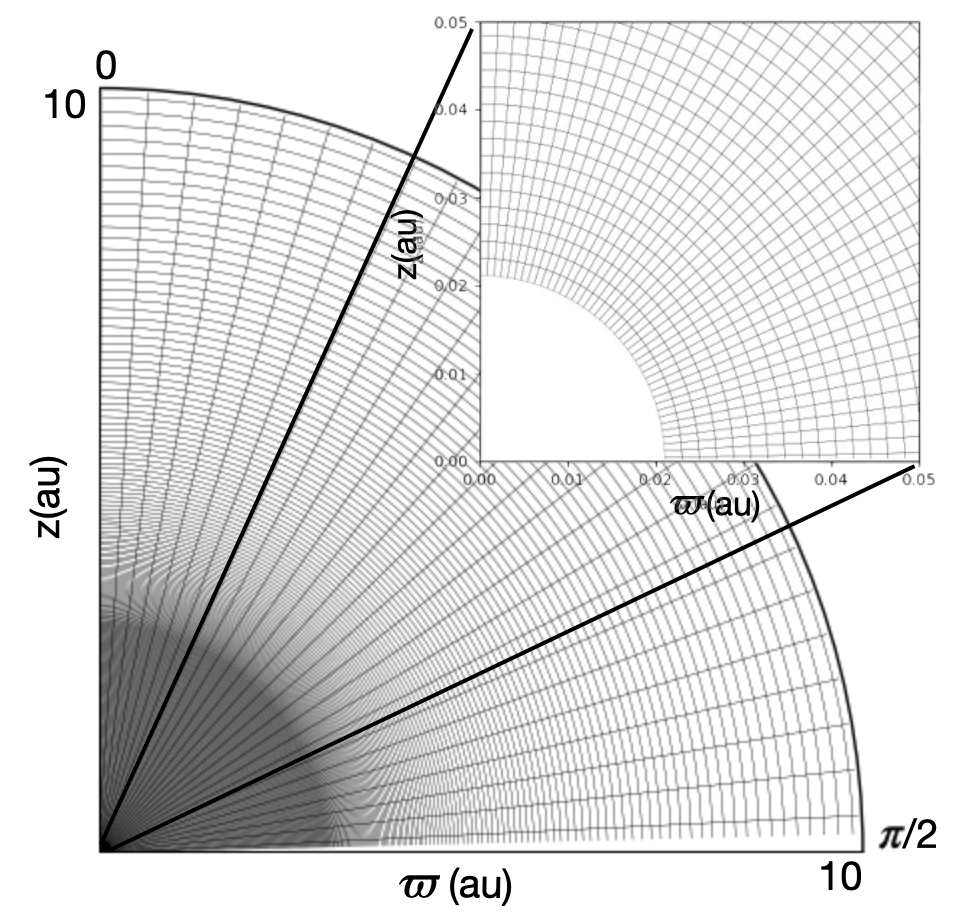}
    \caption{Rarefied computational grid. Proportionality of cells with the original grid is respected. We make sure the grid cells form a square close to the stellar boundary.}
    \label{fig:gridPURE}
\end{figure}

The radial coordinate runs between $r=0.02 \ \rm{au}$ and $r=10 \ \rm{au}$, and has three zones. For $[r_0, r_1]=[0.02, 0.18] \ \rm{au}$ we set a uniform grid of 512 points, then from $r=0.18 \ \rm{au}$ to $r=3 \ \rm{au}$ we use a less accurate grid resolution of 1024 points. Finally, between $r=3 \ \rm{au}$ to $r=10 \ \rm{au}$ we adopt a stretched grid spanning 512 points where the mesh size increases with the radial component (Fig.\ref{fig:gridPURE}). We also run simulations that extend to $30 \ \rm{au}$ that show the existence of a standing recollimation point in  Sec. \ref{sec:standing recollimation}.

The fine grid in the radial direction results in a slightly compressed grid along $r$ (Fig.\ref{fig:gridPURE}). Other less compressed grids were tested, namely by increasing theta to 256 points or decreasing the number of points in the radial direction, but no differences were noted. We obtained higher or lower resolution plots and no change in the dynamics. The run time drastically increased on the other hand. A coarser grid in $R$ smoothes the density resulting in the disappearance of shocks. Overall, changing the grid did not change the steady state reached.

\subsection{Parameters and normalization} \label{sec:normalization}
The ASO solution is determined by four parameters. The ADO solution, on the other hand, has five free parameters. These parameters describe the dynamics of the self-similar models, as well as allow us to normalize all the components to the same stellar object. The parameter set used in this work are all presented in Tab. \ref{ASO_ADO_param}. The parameters $\lambda_d$ and $\lambda_s$ are related to the rotational velocity of the disk wind and stellar solutions, respectively. They provide a measure of the lever arm. Therefore, they also give an indication of the braking efficiency of each solution. The parameters $\delta$ and $\kappa$ control the longitudinal profile of the pressure and density of the ASO solution. The parameters $\mathcal{K}$ and $\nu$ impose the gravitational potential for each solution. $\mu$ is related to the relative magnitudes of magnetic and thermal pressure. The parameter $\gamma$ is the polytropic index. Another parameter is needed to match the two solutions together. It is labeled $\alpha_m$, and defines the isocontour which encloses a constant magnetic flux. More details on its position and role in mixing ASO and ADO solutions are given in Sec.
\ref{mixing proced}. $\mathcal{K}$ and $\nu$ are both expressed in Eq.\ref{nukappa}, as they are directly used to normalize the solutions. The explicit expressions of the other parameters are introduced in Appendix \ref{annex param} when applicable.

\begin{table}[htbp]
\centering
\caption{Parameters used in the ADO and ASO solutions.}
\begin{tabular}{ccccc}
 
 \multicolumn{5}{c}{ADO parameters} \\
 \hline
 $\lambda_d$ &  $\mu$ &$\gamma$ & $\mathcal{K}$ & $x$ \\
 11.7 & 2.99 & 1.05 & 2.0 & 0.75\\
 \hline
 \\
 \multicolumn{5}{c}{ASO parameters} \\
 \hline
 $\lambda_{S}$ & $\delta$ & $\kappa$ & $\nu$ \\
 0.775 & 0.0778 & 0.021 & 1.5\\
 \hline
\end{tabular}
\label{ASO_ADO_param}
\end{table}

To normalize the solutions, three scaling ratios have been chosen at the Alfv\'enic surface. They are nondimensional and denote by a $\star$. Values from the ASO solution are taken as a reference to scale the ADO solution. The scale length, velocity, and density ratios are, respectively,
\begin{equation}
    l_r = \frac{r_{\star d}}{r_{\star s}} , \\ l_v = \frac{V_{\star d}}{V_{\star s}} , \\ l_\rho = \frac{\rho_{\star d}}{\rho_{\star s}} ,
\end{equation}
$r_{\star s}$, $V_{\star s}$, and $\rho_{\star s}$ are all set to unity hereafter. The length factor is set by choosing where the two solutions are mixed together. The position of ADO and ASO reference field lines ($\alpha_d = \alpha_s = 1$, Appendix \ref{annex param}) on the equator gives the normalization factor for length. 
%The explicit expressions of $\alpha_d$ and $\alpha_s$ are in Appendix \ref{annex param}). 
The meeting points of the two solutions are not unique, but the normalization factor is.
We obtain $r_{\star s} = 1$ and $r_{\star d} = 6.25$. The reference field line is located at $\sim 0.1 \ \rm{au}$. Observations indicate that the launching region of disk winds is between $0.2 - 3 \ \rm{au}$ \citep{coffey2004rotation}. By setting up the disk wind closer to the star than $0.2 \ \rm{au}$, we ensured the right physics was in the launching region. From $\mathcal{K}$ and $\nu$ we can determine $V_{\star d}$ and $V_{\star s}$ as both describe the gravitational potential as shown here,
\begin{equation}\label{nukappa}
    \mathcal{K} = \sqrt{\frac{\mathcal{GM}}{\varpi_{\star d} \ V^2_{\star d}}} \\ {\rm{and}} \\
    \nu = \sqrt{\frac{2 \mathcal{GM}}{r_{\star s} \ V^2_{\star s}}}.
\end{equation}
$\mathcal{GM}$ depends only on the star mass. Using this information we get $l_V$ from Eq.\ref{nukappa}, and determine that $V_{\star s} =1$, and $V_{\star d} = 0.212$. The last parameter, $l_\rho$, can be chosen freely. We construct this ratio so as to make the density of the ASO and ADO solutions equal at the equator. The effect of this free parameter will be discussed in Sec.\ref{var rho}. With the density ratio, we impose the magnetic field normalization factor as the two are bound by $V_a = B/\sqrt{4 \pi \rho}$.

In order to return to physical units, we set the velocity, length, and density units from \cite{sautyetal11}. Radius and velocity are normalized at the Alfv\'en radius. Then, the factor for the radius and velocity are $r_0 = r_{alf} = 0.104 \ \rm{au}$ and $V_0 = 106.84 \ \rm{km \ s^{-1}}$, respectively. The density is $\rho_0 = 2.48 \times 10^{-15} \ \rm{g \ cm^{-3}}$. With these units we can subsequently determine pressure density, $P_0 = \rho_0 V_0^2 = 0.28 \ \rm{dyne \ cm^{-2}} $. The magnetic field factor is defined as follows, $B_0 = \sqrt{4\pi P_0} = 1.87 \ G$. Finally, the time scale is, $t_0 = r_0 / V_0 = 1.67$ days.

\subsection{The mixing function} \label{mixing proced}
The semi-analytical solutions that are used in this work have different symmetries and values. This difference introduces discontinuities we want to avoid by smoothly transitioning from one solution to the other in the numerical grid. Using a mixing function to link two objects consistently is not new \citep{Matsakosetal2009,Matsakosetal2012} and is still currently used. For instance, \cite{jannaud2023numerical} have used a spline function that depends on $\theta$ to go from a nonrotating central object to a Keplerian disk at the boundary. In \citep{Matsakosetal2009,Matsakosetal2012}, the mixing function is in terms of the magnetic flux, and used on all physical quantities that describe the evolution of the plasma. 

The mixing procedure in this work is inspired from \cite{Matsakosetal2009}. We adapt their mixing procedure but we keep the launching region close to the star close to the initial setup  presented in \cite{sauty2022nonradial}. It ensures a better treatment of the inner accretion zone and makes it easier to compare the various simulation results. Additionally, by refining the mixing function presented in \cite{Matsakosetal2009} we chose to reduce the mixing parameters from two free parameters to only one, which controls the mixing width.

We use the magnetic flux as the main mixing parameter. The purpose of which is to avoid introducing regions where $\nabla \cdot {\bf B} \neq 0$ at the initial step. $\mathbf{B}$ depends on the magnetic flux $A(\alpha)$ as shown in Eq.\ref{Bcalc}. Given $A(\alpha)$ has a different slope for each component, special care has to be considered to ensure a divergence-free magnetic field at the boundary conditions,
\begin{equation}\label{Bcalc}
    B_p = \frac{\nabla A \times \hat{\phi}}{\varpi}.
\end{equation}
We define the mixing function as follows,
\begin{equation}\label{mixing function}
    U = w_s U_s + w_d U_d
\end{equation}
with
\begin{equation} \label{mixing param}
    w_s = exp\left[-\left(\frac{\alpha_{s}}{\alpha_{m}}-1\right)^d\right] \\ \rm{and} \\ w_d = 1 - w_s,
\end{equation}
$\alpha_{s}$ is the nondimensional magnetic flux of the stellar solution and $\alpha_{m}$ is a constant defining the last open field.
Along the polar axis, $w_s$ is equal to one until the last open fieldline. Then, starting the last open field, it decreases exponentially to zero as the cylindrical radius increases. 

\begin{figure}[htbp]
        \centering
    \includegraphics[width=0.45\textwidth]{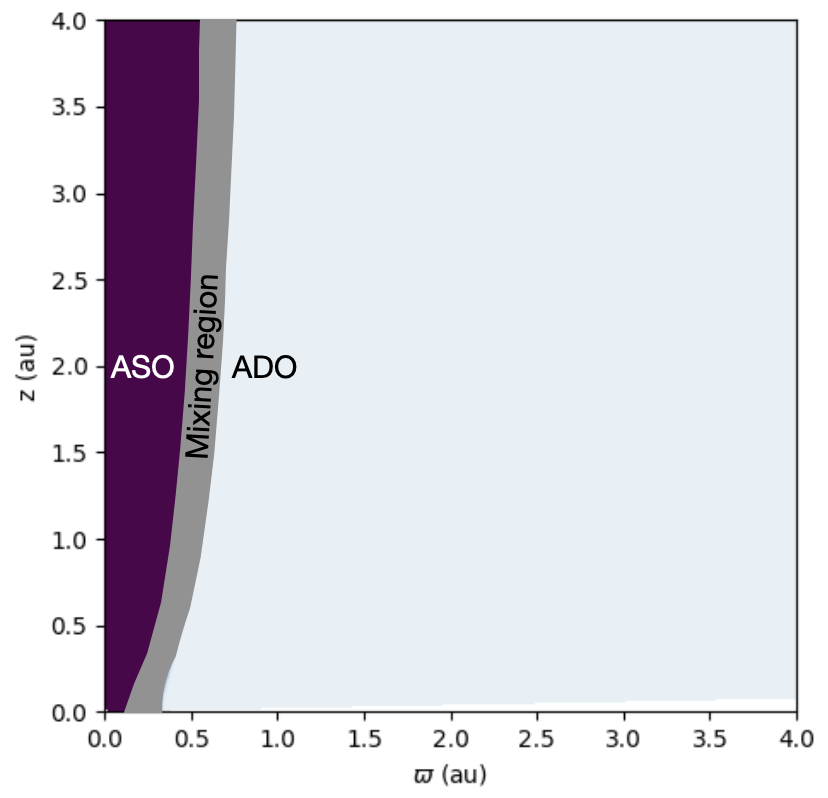}
    
    (a)
        
     \includegraphics[width=0.45\textwidth]{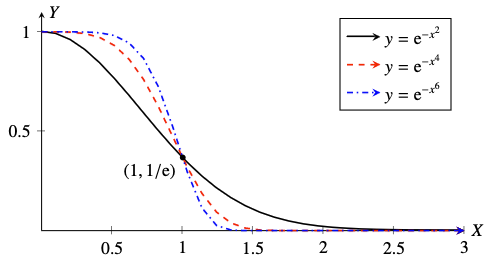}
    
    (b)
    
        \caption{(a) the 2D map of the mixing function that goes from zero (Blueish region) to one (purple region). $\varpi$ is the cylindrical radius and $z$ the rotational axis. The jet propagates along the $z$ axis and $\varpi$ is the cylindrical radius in the equatorial plane. A gray mask was added to better denote the mixing region. 
 %\textcolor{magenta}{CS then the color bar is useless? How does it look with no grey?} 
    The speed of transition inside this region can be controlled through $d_m$, as shown in (b). (b) illustration of the extension of the ASO component over the selected region. The functions are of qualitative value. This shows the dependence of the parameter d on the mixing speed. We also shift the center of the function in this plot to zero. This allows for a better presentation of the mixing parameters.}
        \label{fig:dm behavior}
\end{figure}

The last open magnetic field, associated in the stellar solution to the iso-contour $\alpha_m \sim 1$, sets the strip along which the mixing occurs (gray region in Fig.\ref{fig:dm behavior}). The width of the mixing, that is, the speed at which we transition from the stellar solution to the disk solution, is controlled by the parameter $d_m$ in Eq.\ref{mixing param}. We can see in Fig.\ref{fig:dm behavior} that the higher $d_m$ is chosen, the steeper the transition becomes. Another consequence of increasing $d_m$ is expanding the region where $w_s$ is equal to one. In fact, this has the same effect as pushing the ASO component further into the disk wind by increasing $\alpha_m$. Given this behavior, we fix $d_m=2$ to insure $w_s$ starts decreasing soon after entering the ADO region. The wind disk should have a higher density than the ASO solution can provide in the transition region.

The physical quantities $\rho$, $P$, $V_r$, $V_\phi$, $B_\phi$, and $A$, are all blended using the mixing function presented above. The exception to this rule are $\mathbf{B_p}$ and $\mathbf{V_p}$. The poloidal components of the magnetic field are computed with Eq.\ref{Bcalc}, which takes into account derivatives of the weights. Then, in ideal MHD, the poloidal velocity is parallel to the poloidal magnetic field. Thus, the velocity can be expressed as,
\begin{equation} \label{VB}
     V_\theta = V_r \frac{B_\theta}{B_r}.
\end{equation}
where $V_r$ is computed as shown in Eq.\ref{mixing function}.

\subsection{The heating function} \label{sec: heating func}
From photoevaporation to viscous heating, an abundant number of processes can explain how the disk and its environment may be heated, but the details of such processes are still not resolved. We opt to consider two different atmospheres associated to the spine outflow and the disk jet, as there is no reason for the two regions to have the same physical origin, order of magnitude in terms of heating, or even the same geometry. As a first approach to the problem, we use the analytical heating provided by the analytical solutions, and introduce parameters that will help us control the heating input by parametrizing it.

\begin{figure}[htbp]
    \centering
    \includegraphics[width=0.48\textwidth]{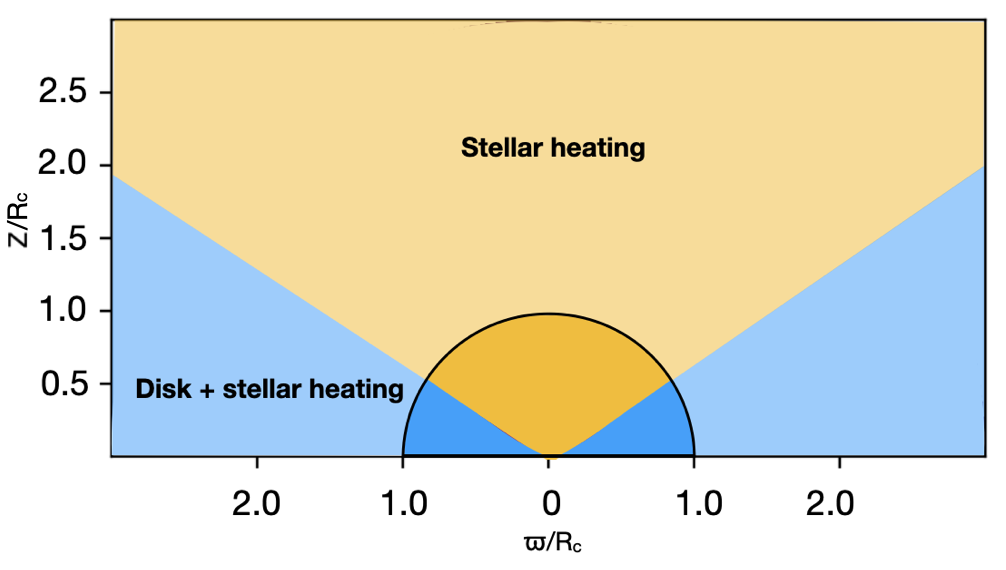}
    \caption{Scheme of heating allocation to the stellar and disk wind components. The black half circle indicates the spherical nature of the parameter $R_c$. The fading colors refer to where the heating is suppressed slowly using the method presented in Eq.\ref{chauffweightASO}. The heating profiles along the polar and equatorial axes are given in Fig. \ref{fig:heating v radius}.
    % NV suggestion: 2D PLOT of heating map normalized to rho V E/varpi 
    }
    \label{fig:comp heating}
\end{figure}

The disk wind uses the values of Eq.\ref{ADo heating} (ADO heating) mixed with Eq.\ref{ASO heating} (ASO heating) as decribed in the mixing procedure.
%The ADO heating is negligible when the disk wind is alone. But when you are in the mixed region the disk heating is too high compared to the ASO heating (despite the fact that it is negliglible compared to the magnetic energy) and it destroys the simulation.
Heating above the disk is contained in a smaller region of the disk component. The reason is that velocities are too high, which leads Eq.\ref{ADo heating} to produce non physical heating rates. Given the physical limitations linked to this equation, we restrain the disk atmosphere from the equator to around $20^{\circ}$ above the disk. We use the stellar heating equation (Eq.\ref{ASO heating}) for the rest of the domain, that is, part of the disk and spine jet components (orange part of Fig.\ref{fig:comp heating}). 

We further introduce a decreasing function to force the heating rate of both the disk and the spine outflow to go toward zero beyond a radius we call $R_c$. This radius is a free parameter that controls where the heating starts to smoothly dip. We also introduce a second free parameter, $I_h$, which in turn controls the heating in the disk atmosphere. Eq.\ref{chauffweightASO} describes the final form of the total heating the whole simulation box,
\begin{equation}\label{chauffweightASO}
Q (r) = \left\{
    \begin{array}{ll}
        Q_{an} \times \arctan (\frac{1}{(r-R_c)^4}) & \mbox{orange region} \\
        I_h \times Q_{an} \times \arctan (\frac{1}{(r-R_c)^4}) & \mbox{blue region}
    \end{array}
\right.
,\end{equation}
where $Q_{an}$ is the analytical heating computed with the solutions pressures and velocities dispatched as shown in Fig.\ref{fig:comp heating}.  $Q_{an}$ is computed using Eq.\ref{ASO heating} in the stellar atmosphere (orange region in Fig.\ref{fig:comp heating}). In the disk atmosphere (blue region in Fig.\ref{fig:comp heating}), $Q_{an}$ is computed as follows:
\begin{dmath}
    Q_{an} = w_s \times \rho_{ ASO} {\bf V_{ ASO}} \cdot \left[\Gamma \nabla \left(\frac{P_{ ASO}}{\rho_{ ASO}}\right) - (\Gamma-1) \frac{\nabla P_{ ASO}}{\rho_{ ASO}}\right] + w_d \times (\Gamma - \gamma)P_{ ADO}(\nabla \cdot \mathbf{V_{ ADO}}).
\end{dmath}
\noindent In Sec.\ref{sec: ParamSpace} we present the behavior of the outflow over the parameter space.

\subsection{Boundary conditions}
At the outer radial boundary, the quantities that describe the plasma are fixed to the analytical solutions values. If on the contrary we set the gradient, on every physical quantity, to be zero at the boundary by imposing outflow conditions, we note that the system develop over-density waves created below the Alfv\'en surface which bounce back toward the jet. These waves artificially modify the jet collimation radius making it narrower. Choosing to impose the analytical solution for all quantities on the boundary does not change the evolution in the super-Alfv\'enic region. However, it dampens the density that reaches the boundary, which in turn prevents the over-density waves from traveling back toward the axis. Consequently, the jet as a whole does not recollimate. \cite{stute2008stability} noticed that imposing an "outflow" condition at the boundary precipitate the collapse of the simulated jets. This collapse is thought to be artificially triggered, since all gradients at the outer radial boundary are equal to zero. The terms that survive are related to the pinching force (i.e., $B_\phi/\varpi$) which is directed toward the axis. By damping $B_\phi$ to zero on the outer boundary, the artificial collimation reportedly disappears, which reinforces the conclusion that such an artifact is solely numerical (see model ER2 in \cite{stute2008stability}). Choosing a bigger box only postpones the boundary effect to later times.

\begin{figure}[ht]
    \centering
    \includegraphics[width=0.49\textwidth]{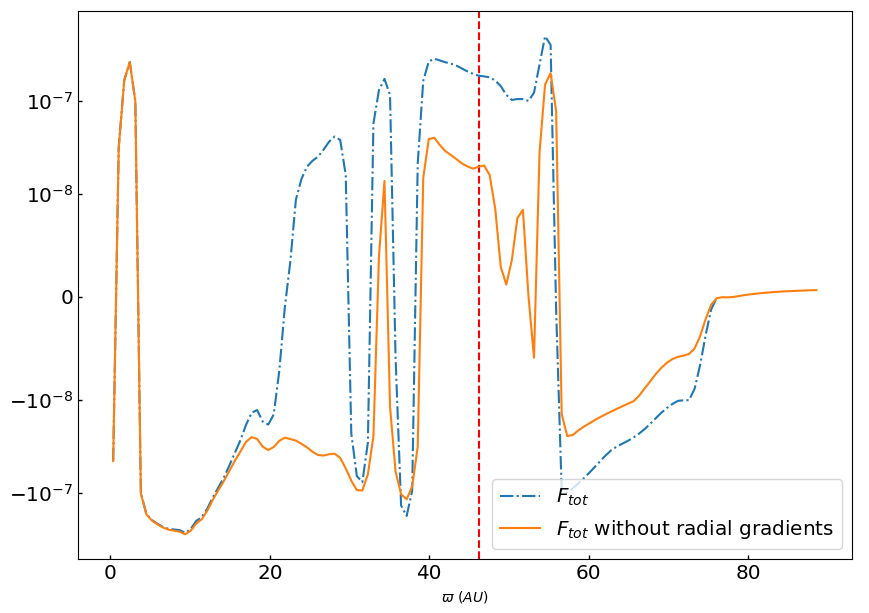}
    \caption{Total force applied by the pressure gradient, gravity, and Lorentz force on the plasma with the radial gradient (blue curve) and then without it (orange curve). We fix R at the radial boundary and vary the colatitude $\theta$. The vertical line separates the super-Alfv\'enic and sub-Alfv\'enic regions. The total Lorentz force density is computed using $\bf{J} \times \bf{B}$ adapted to a spherical grid.}
    \label{fig:Force w/,w/out radial comp}
\end{figure}

To identify the cause of the jet narrowing, we plot the Lorentz force density from grids with analytical boundary conditions and outflow conditions. In Fig.\ref{fig:Force w/,w/out radial comp}, we have a representative example of what we get.  
We see a difference between the total force if the radial gradient is included or not. This is the case for all simulations, but no clear behavior emerges. We argue that the radial boundary in our case acts as a wall that partially reflects matter. Since the reflection happens in the sub-Alf\'enic region, the wave travels back and changes the jet dynamics. To remove the reflection we set the outer radial boundary to always have the values defined by the analytical solution. By doing so, we mitigate any over-density that reach the boundary. Subsequently, artificial boundary effects are suppressed.  
\cite{stute2008stability} on the other hand dampen $B_\phi$, only, to zero.

In summary, it is crucial to either minimize the toroidal magnetic field component $B_\phi$ at the outer radial boundary or make self-consistent modifications to all quantities, ensuring equilibrium in the external region.

\section{Results} \label{sec:results}

%{\bf AM: un commentaire general pour CS aussi: je pose la question : certes on met en évidence une variation de morphologie des jets suivant l'effet de deux paramètres. Mais a-t-on des arguments pour dire que ce sont vraisemblablement les deux principaux paramètres qui contrôlent cette morphologie... pour le moment de ce que j'ai lu ce n'est pas évident.
%\textcolor{magenta}{CS:  AM is right. Some how the two heating parameters controls the dynamics and the morphology as long as the main collimation parameter is the magnetic field. So it is not the heating per se but comparing to the effects of the magnetic field.}

%\textcolor{magenta}{CS: when you make changes to answer my comments, it would be nice to put highlight it with another color like red or blue}

\subsection{Parameter space} \label{sec: ParamSpace}
Our study concentrates on the effect of our two main controlling parameters. The first parameter $I_h$ controls the intensity of the heating computed from the disk. As shown in Eq.\ref{chauffweightASO}, the heating is computed from the disk, then multiplied by $I_h$ to lower the heating rate input. The second parameter $R_c$ sets the radius at which heating is smoothly dropped in the computational grid. The first parameter is a simple constant we append to the disk heating rate. The purpose of changing the heating intensity is to gauge how the heating affects the magnetic collimation of the jet, given the intensity of heating of the disk wind. The second parameter allows one to evaluate how small the heating area around the YSO can be for the jet to be maintained.

\begin{figure}[ht]
    \centering
    \includegraphics[width=0.48\textwidth]{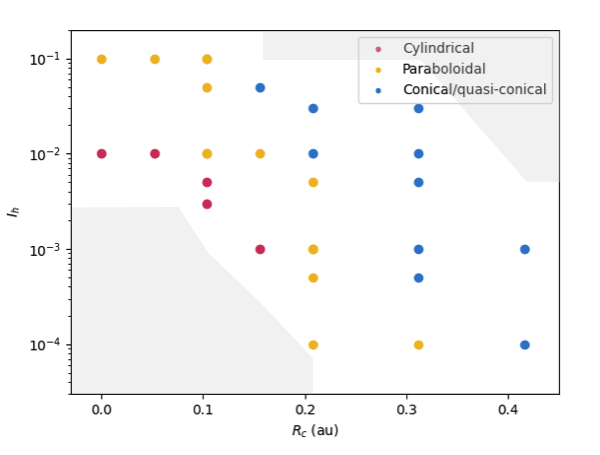}
    \caption{Parameter space mapping the simulations performed. Each dot is a set of parameters. Red dots are cylindrical jets, yellow is for paraboloidal jets, finally the blue dots represent wide paraboloidal or conical solutions. The gray areas are parameter space zones not covered by MHD simulations because the code crashes systematically.}
    \label{fig:param space}
\end{figure}

\begin{figure*}[htbp]
    \centering
    {\color{red} (a) Cylindrical}
    \\
    \includegraphics[width=.54\textwidth]{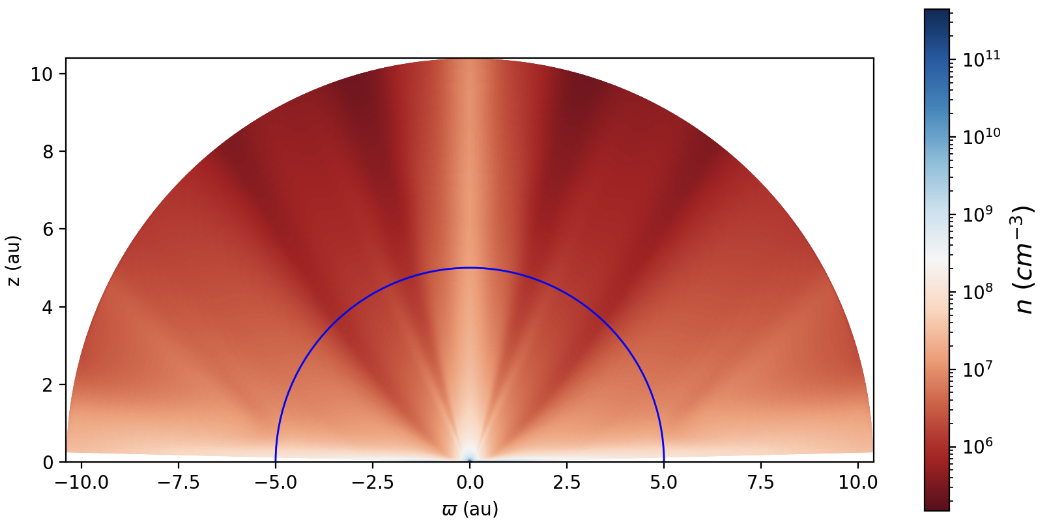} 
    \includegraphics[width=.4\textwidth]{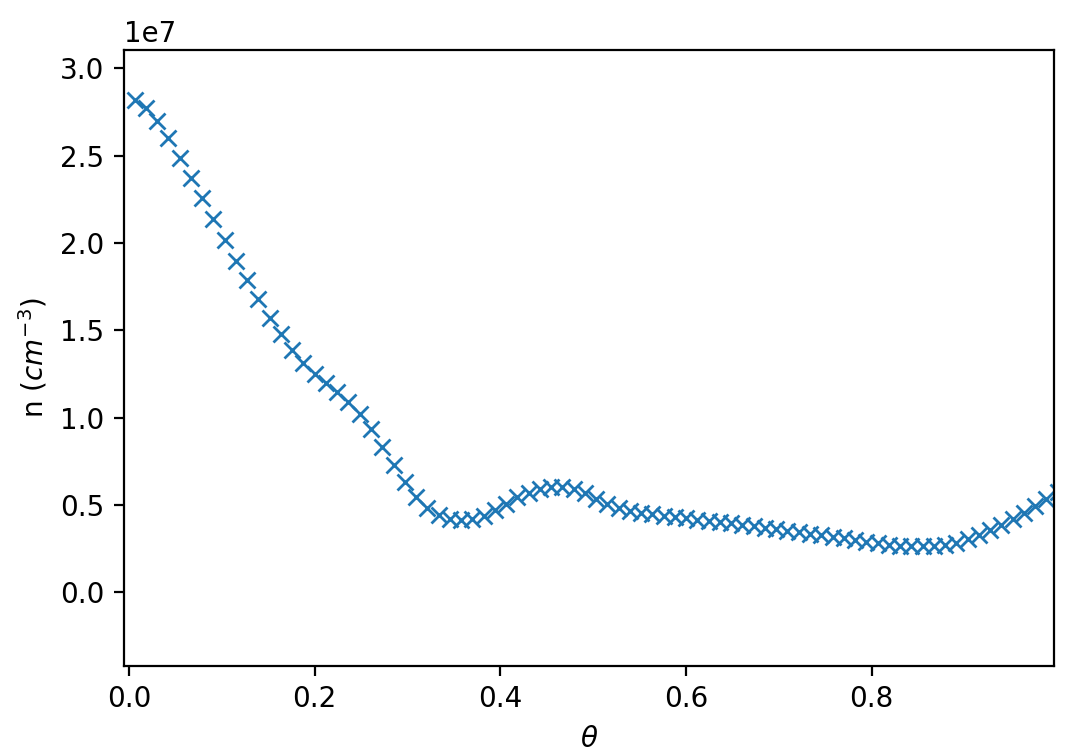}
    
    {\color{yellow} (b) Paraboloidal}
    \\
    \includegraphics[width=.54\textwidth]{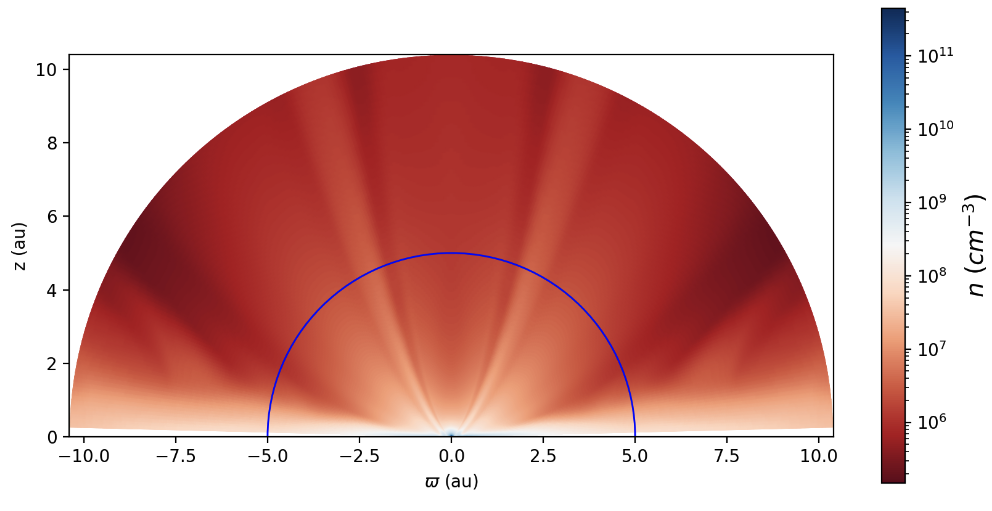}
    \includegraphics[width=.4\textwidth]{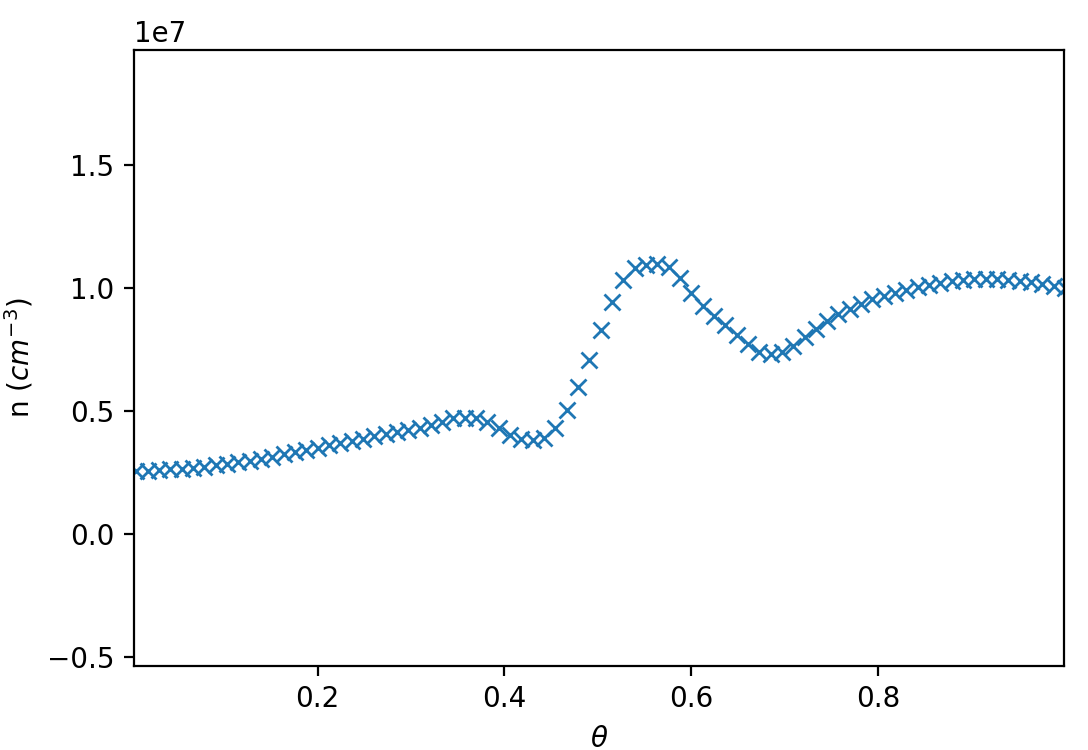}
    
    {\color{blue} (c) quasi-conical}
    \\
    \includegraphics[width=.54\textwidth]{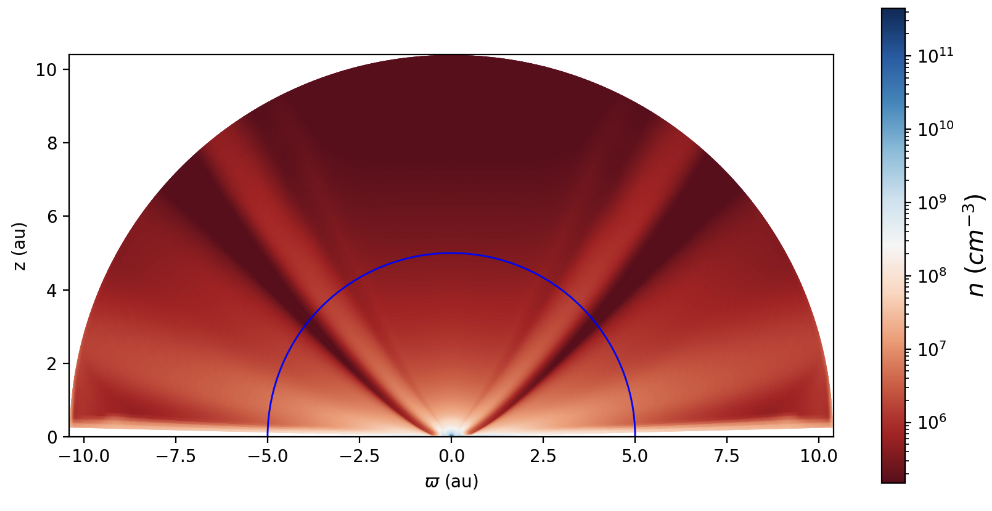}
    \includegraphics[width=.4\textwidth]{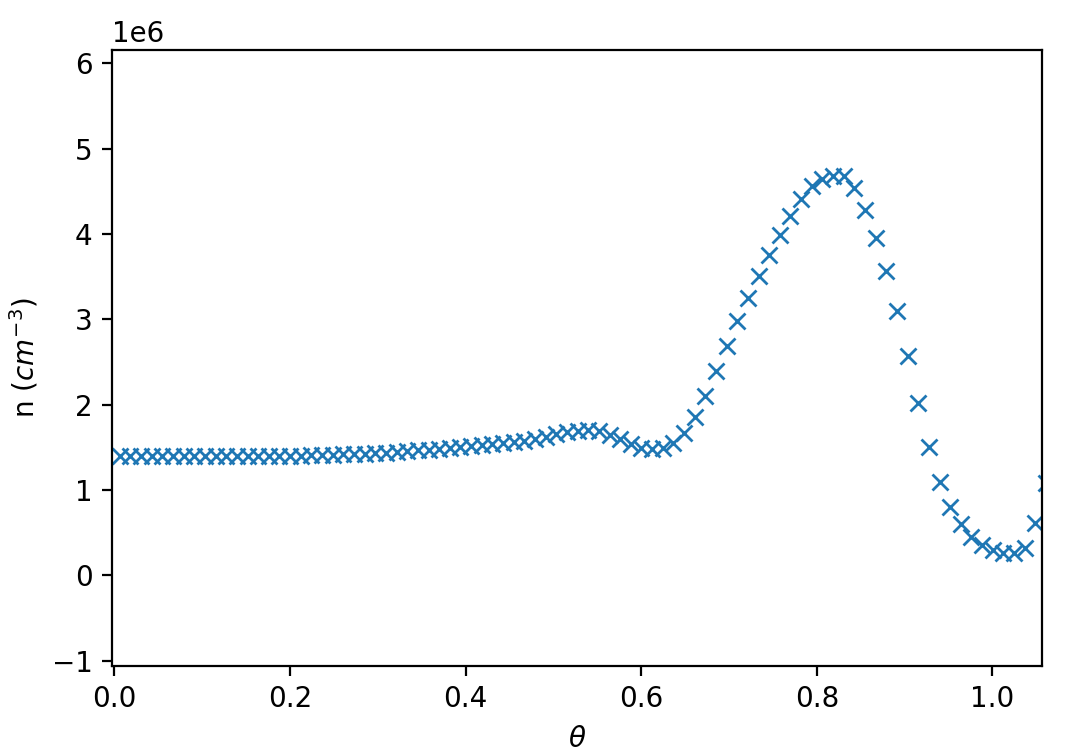}
    \caption{Plots of jet shapes found in the parameter space. The 2D maps of density show the collimation up to $10 \ \rm{au}$. What differs between {\color{red} cylindrical} and {\color{yellow} paraboloidal}, in this case, is the density at the axis. Generally, the opening angle is sufficient to distinguish {\color{red} cylindrical} from {\color{yellow} paraboloidal} solutions. The numerical diffusion is so low that the density isocontours coincide with the magnetic flux tubes and the poloidal velocity line (as in Fig. \ref{fig:ME production}). The more collimated the flow is, the lower the acceleration efficiency.  When this does not apply we look at the density distribution on the polar axis. For (a), $R_c = 0 \ \rm{au}$ and $I_h = 10^{-1}$ (R0\_I1). Next, for (b), $R_c = 0.2 \ \rm{au}$ and $I_h = 10^{-3}$ (R2\_I3). Finally, for (c), $R_c = 0.4 \ \rm{au}$ and $I_h = 10^{-3}$ (R4\_I3). The blues half circle on all the cases indicates the radius along which the number density is plotted. We plot the density along $\theta$ starting at the axis and stop before reaching the disk wind region.
    % PLOT M_f above the density along the same circle. 
    }
    \label{fig:morphs jet}
\end{figure*}

\begin{figure}[ht]
    \centering
    \includegraphics[width=.445\textwidth]{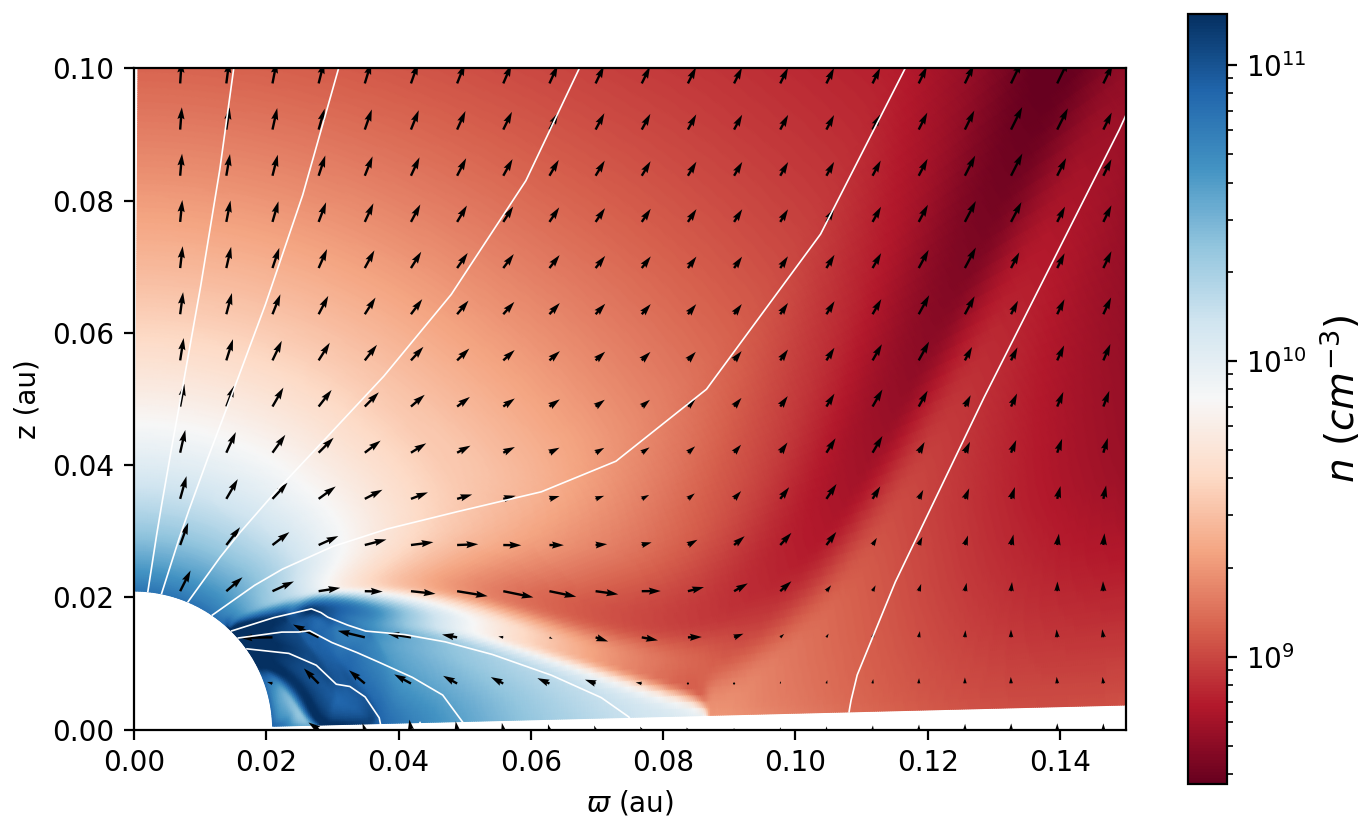}
    \includegraphics[width=.445\textwidth]{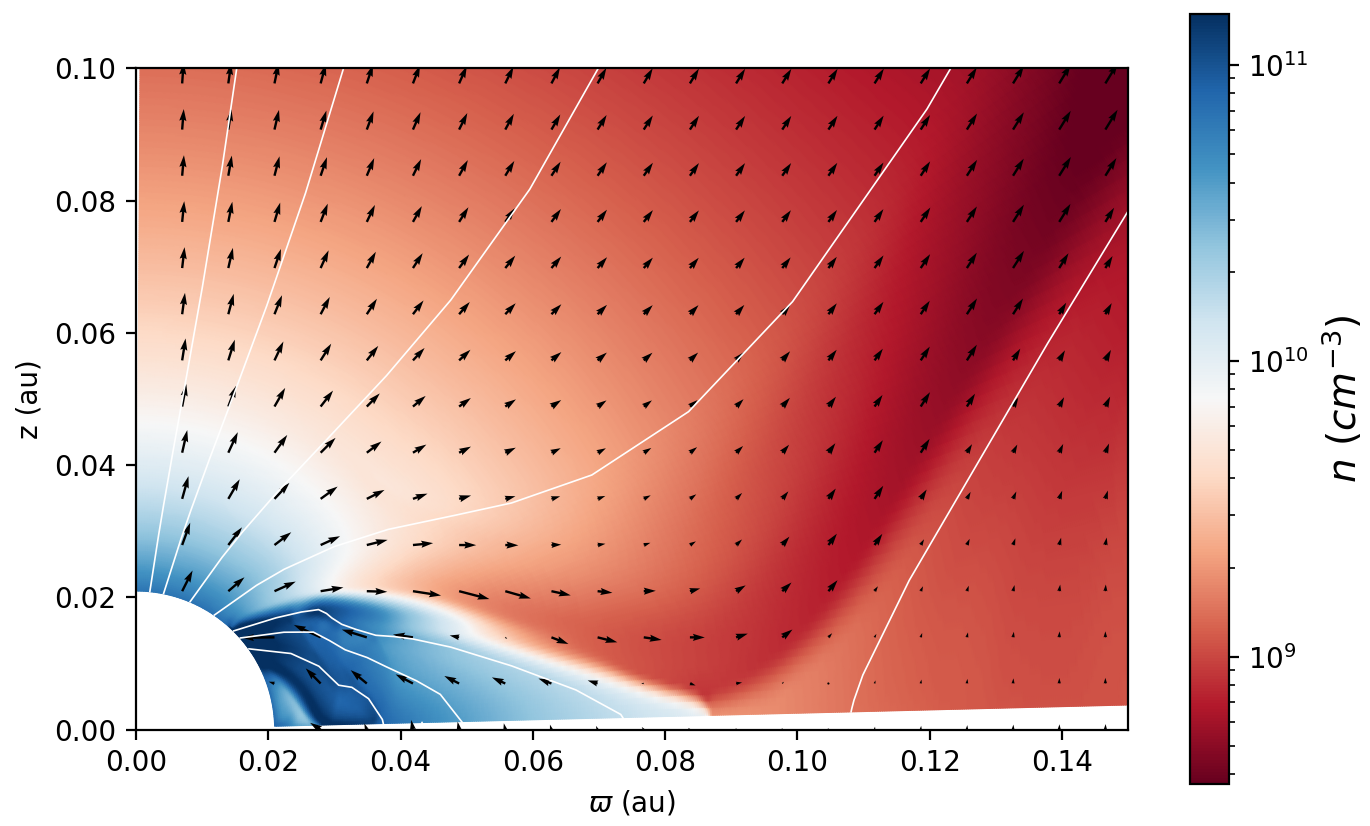}
    \includegraphics[width=.445\textwidth]{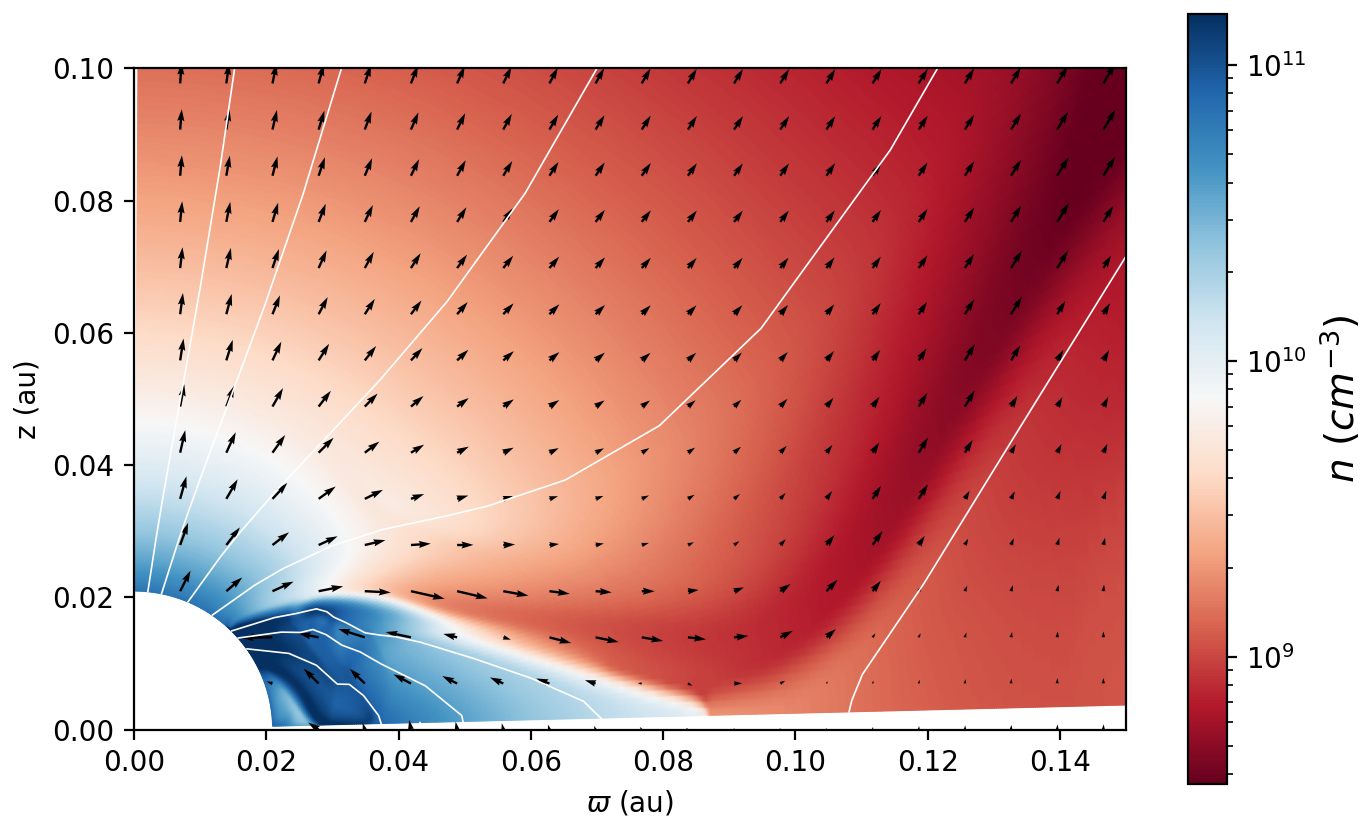}
    \includegraphics[width=.445\textwidth]{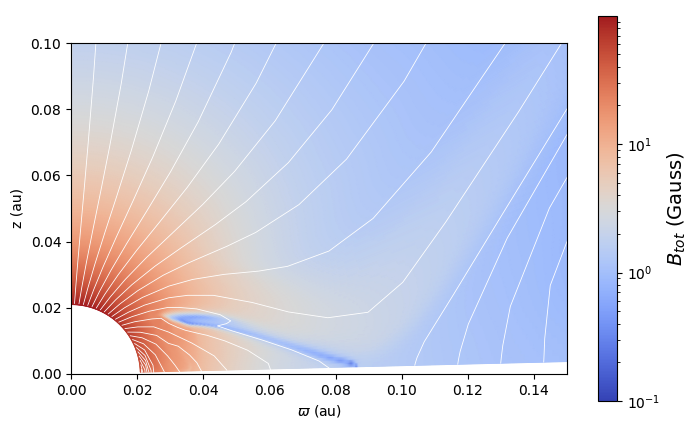}
    \caption{Close-up  to the star for the three cases shown in Fig.\ref{fig:morphs jet}. Although the envelope changes dramatically between morphologies, the central component is similar through out the cases. The denser region with magnetic fields connecting the star and the disk boundary is the accretion region. The last plot is of the total magnetic field 2D profile close to the star.}
    \label{fig:close up star}
\end{figure}

Fig.\ref{fig:param space} presents the results produced by the parameter variations. We obtain three basic morphologies with a smooth transition between the different cases. We distinguish each case using tracers that follow the spine jet, and part of the disk material emanating from radii $\sim 0.9 - 3 \ \rm{au}$. The semi-opening angle of the spine jet is then determined close to the radial boundary condition ($10 \ \rm{au}$). For the narrowest spine jet (cylindrical solution), the semi-opening angle is $<10^\circ$, the widest configurations (quasi-conical/conical solutions) have semi-opening angles $>20^\circ$, and in between lays the paraboloidal solutions. An additional way of distinguishing cylindrical and paraboloidal solutions is by looking at the density along the axis. All the narrow cylindrical morphologies have a dense spine outflow with most of the density concentrated around the polar axis, as can be seen in Fig.\ref{fig:morphs jet}. The close-up to the central engine is displayed in Fig.\ref{fig:close up star}. This eases the distinction between cylindrical and paraboloidal solutions regardless if the opening angles overlap. Finally, while for lower $I_h$ the jet slowly transitions from one morphology to the other, this change is faster for $I_h \sim 10^{-2}$ then becomes slow again for higher $I_h$. The gray regions in Fig.\ref{fig:param space} represent where simulations crash.

Our simulations do not contain a lower density threshold as it is common to implement in MHD simulations. Because of this choice, the space parameter is void of points in the top right corner in Fig.\ref{fig:param space}. As we heat up matter, the density in regions close to the equator drops drastically.
Since MHD codes have a weakness treating matter-poor environments, this leads to a lack of points. The lower left corner, on the other hand, does not contain points, because without sufficient heating, matter accretes and creates a strong gradient against the inner region close to the star. The code time increment decreases drastically as a consequence, making it impossible to explore these areas.
Both gray regions in Fig.\ref{fig:param space} are of no major interest to us since the jets produced are either too wide (upper right region) or too narrow (lower left region). Thus, we choose to not add a density threshold to keep artificial effects to a minimum, and maintain coherence between simulations. 

Since for all the simulations presented here the magnetic field distribution is constant, adding more heating is equivalent to enhancing the pressure gradient. The additional pressure is located outside the spine jet and above the disk. As we heat up this region, the pressure gradient steepens and forces the jet to open up more.

\begin{table*}[ht]
    \centering
    \caption{List of the simulations performed with analytical radial boundary conditions. Except for R0\_I2 and R05\_I2, all simulations reach a steady state. The heating rate is computed by considering its mean value at the disk atmosphere up to $1 \ \rm{au}$. All the values presented here are extracted at the final time of the simulations. $R_{JED}$ is the last radius at which the disk launches a jet.}
    \bgroup
    \def\arraystretch{1.35}
    \begin{tabular}{lcccccc}
    \hline
    \hline
         Sim ID & $R_c \ (\rm{au})$ & $I_h$ & Avrg. Disk Heating  ($\rm{erg \ cm^{-3} \ s^{-1}}$) & $\dot{M}_{JED} \ (\Msunyr)$ & $\dot{M}_{stel} / \dot{M}_{tot}$ & $R_{JED}$ ($\rm{au}$)\\ 
         \hline
         R0\_I1 & $0$ & $10^{-1}$ & $6.2 \cdot 10^{-7}$ & $6.8 \cdot 10^{-9}$ & 0.31 & 0.42\\ 
         R0\_I2 & $0$ & $10^{-2}$  & $1.7 \cdot 10^{-7}$ & not steady & only stellar - 0.15 & no JED - 3 au \\ 
         R05\_I1 & $0.052$ & $10^{-1}$  & $1.8 \cdot 10^{-6}$ & $1.2 \cdot 10^{-9}$ & 0.69 & 0.21\\ 
         R05\_I2 & $0.052$ & $10^{-2}$  & $2.9 \cdot 10^{-7}$ & not steady & only stellar - 0.23 & no JED - 1 au\\ 
         R1\_I1 & $0.104$ & $10^{-1}$  & $5.6 \cdot 10^{-6}$ & $6.5 \cdot 10^{-10}$ & 0.82 & 0.19 \\
         R1\_I12 & $0.104$ & $0.5 \ 10^{-1}$  & $2.9 \cdot 10^{-6}$ & $1.3 \cdot 10^{-9}$ & 0.74 & 0.21\\ 
         R1\_I2 & $0.104$ & $10^{-2}$  & $5.4 \cdot 10^{-7}$ & $7.5 \cdot 10^{-9}$ & 0.24 & 0.63\\ 
         R1\_I22 & $0.104$ & $0.5 \ 10^{-2}$  & $4.7 \cdot 10^{-7}$ & $9.7 \cdot 10^{-9}$ & 0.21 & 0.63\\ 
         R1\_I23 & $0.104$ & $0.3 \ 10^{-2}$  & $3.7 \cdot 10^{-7}$ & $1.1 \cdot 10^{-8}$ & 0.21 & 0.63\\
         R12\_I12 & $0.156$ & $0.5 \ 10^{-1}$  & $3.6 \cdot 10^{-6}$ & $6.4 \cdot 10^{-10}$ & 4.6 & 0.19\\
         R12\_I2 & $0.156$ & $10^{-2}$  & $8.7 \cdot 10^{-7}$ & $1.5 \cdot 10^{-9}$ & 0.51 & 0.20\\
         R12\_I3 & $0.156$ & $10^{-3}$  & $2.6 \cdot 10^{-7}$ & $1.0 \cdot 10^{-8}$ & 0.29 & 0.60\\
         R2\_I13 & $0.208$ & $0.3 \ 10^{-1}$  & $2.8 \cdot 10^{-6}$ & $1.1 \cdot 10^{-9}$ & 0.81 & 0.20\\ 
         R2\_I2 & $0.208$ & $10^{-2}$  & $1.0 \cdot 10^{-6}$ & $2.3 \cdot 10^{-9}$ & 0.61 & 0.26\\ 
         R2\_I22 & $0.208$ & $0.5 \cdot 10^{-2}$  & $5.9 \cdot 10^{-7}$ & $3.5 \cdot 10^{-9}$ & 0.42 & 0.32\\
         R2\_I3 & $0.208$ & $10^{-3}$  & $2.2 \cdot 10^{-7}$ & $4.6 \cdot 10^{-9}$ & 0.28 & 0.37\\
         R2\_I32 & $0.208$ & $0.5 \cdot 10^{-3}$  & $1.7 \cdot 10^{-7}$ & $6.2 \cdot 10^{-9}$ & 0.24 & 0.41\\
         R3\_I13 & $0.312$ & $0.33 \cdot 10^{-1}$  & $3.6 \cdot 10^{-6}$ & $7.0 \cdot 10^{-10}$ & 0.85 & 0.19\\
         R3\_I2 & $0.312$ & $10^{-2}$  & $1.6 \cdot 10^{-6}$ & $2.0 \cdot 10^{-9}$ & 0.66 & 0.25\\
         R3\_I22 & $0.312$ & $0.5 \cdot 10^{-2}$  & $7.0 \cdot 10^{-7}$ & $3.4 \cdot 10^{-9}$ & 0.54 & 0.28\\
         R3\_I3 & $0.312$ & $10^{-3}$  & $2.7 \cdot 10^{-7}$ & $5.5 \cdot 10^{-9}$ & 0.34 & 0.36\\ 
         R3\_I32 & $0.312$ & $0.5 \cdot 10^{-3}$  & $2.5 \cdot 10^{-7}$ & $6.5 \cdot 10^{-9}$ & 0.29 & 0.48\\
         R3\_I4 & $0.312$ & $10^{-4}$  & $1.4 \cdot 10^{-7}$ & $9.6 \cdot 10^{-9}$ & 0.23 & 0.52\\
         R4\_I3 & $0.416$ & $10^{-3}$  & $3.2 \cdot 10^{-7}$ & $5.7 \cdot 10^{-9}$ & 0.35 & 0.35\\
         R4\_I4 & $0.416$ & $10^{-4}$  & $1.5 \cdot 10^{-7}$ & $8.0 \cdot 10^{-9}$ & 0.26 & 0.42\\ 
         \hline
    \end{tabular}
    \egroup
    \label{tab:simID}
\end{table*}

In Tab.\ref{tab:simID} we show a mean value of the heating rate taken from the disk atmosphere. We average the heating until $1 \ \rm{au}$. the disk jet mass loss is determined once we exit the spine jet, between two minima that contain a local maximum in density, as shown in fig.\ref{fig:morphs jet}. As for the radius in which ejection from the disk occurs, reported in the last column of the table, we obtain its value by equating the mass loss in the disk jet and at the base of the jet. Following MHD arguments, plasma is not free to cross magnetic flux tubes. Thus, by tracking the flux tube containing the totality of the disk jet mass loss to the base, we deduce the radius of ejection.    

%{\bf AM: pour le moment pour moi tu ne discutes pas la figure 4 pourquoi as tu une solution dans un sous espace donné ? on en sait toujours pas après la lecture l'interpétation physique du phenomène}

\subsection{Varying the disk wind density}\label{var rho}
First and foremost, we have to quantify the effect the disk density has on the dynamics of the jet. To do so, we vary $l_\rho$ and adapt the normalization of the magnetic field to be consistent with the MHD formalism. Tab.\ref{tab:den norm} displays the chosen values for $l_\rho$, $l_B$, and $d_m$. 
When the disk wind has a lower density than the magnetosphere
the disk is not magnetically strong enough to maintain a jet. On the other hand too large densities need a high heating to produce a jet. To constrain the normalization, we use $ProDIMo$, PROtostellar DIsk MOdel, \citep{woitke2009radiation,kamp2010radiation,woitke2016consistent,rab2018x} to determine the surface density of the accretion disk at the jet launching region in the disk. $ProDIMo$ is a thermo-chemical radiation code, where wavelength-dependent radiative transfer calculation are conducted. $ProDIMo$ computes gas and dust temperatures, in addition to the local radiation field. The code, then, computes the chemical abundances and determine the thermo-chemical disk structure taking into account the radiative transfer of an external X-ray flux. The temperature is dictated by the heating and cooling processes based on chemical abundances. This involves a network of 235 different species and 3143 reactions \citep{kamp2017consistent,rab2017stellar}. Due to the uncertainty of the chemical composition and physical surface of such a disk, we find values for disk density between $10^9 - 10^{10} \ cm^{-3}$ at $R = 0.5 \ \rm{au}$ and for $z/r=[0.1,0.2]$. We choose to fix $l_\rho$ at $0.014$ which lays in the density limits inferred from $ProDIMo$. This density is consistent with MHD winds extracted from a standard disk. We will discuss how changing the density modifies our results in this section. 

The solutions for the parameter set $(R_c, I_h)$ displayed in Fig.\ref{fig:param space} is not unique and depends on the disk density. As for the evolution of the space taken by each case in the space parameter, we would have to run another set of simulations changing density. We choose to fix heating and vary density between the maximum and minimum values given by $ProDIMo$. We have run three types of simulations. The first  one uses the maximum value of the density. The second is performed with an intermediary density. The third one is conducted with the minimum density as shown in Tab.\ref{tab:den norm}.

We find that raising the density shifts the parameter space of the jet solutions to the right, while lower density simulations produce higher velocities than what is observed for these objects. We discard the latter configuration, because the space parameter becomes difficult to pave. We describe in the next section the normalization chosen for our parametric study. Hereafter, however we will not further study the effect of density on the parameter space as we reserve it for future work where an other type of disk solutions will be used, such as the solutions presented in \cite{ferreira1997magnetically}. These solutions have a higher mass loss rate and have been upgraded to include a magnetic pressure component that does not hide its contribution behind an effective heating.
%{\bf AM: je ne comprends ce que tu fais: tu discutes les effets d'augmenter la densité ci-dessus, donc cette translation de solutions devrait déjà être discutéee au dessus, tu parles de heating en dessous dans ce paragraphe, mais ça c'est donné par la figure ...  

However, we stress that the present study is, nonetheless, robust. The points associated to each class of solution may shift but the position of the colors stay fixed in relation to each other. The cylindrical jets, then the hyperbolic, and finally the quasi-conical jets all appear in this sequence. The more we heat up the disk wind, the shorter the cross over between colors. We give each dot a name as presented in Tab.\ref{tab:simID} to facilitate the transition between the parameter space and the jet dynamic.
%}
%\begin{figure}[ht]
%    \centering
%    \includegraphics[width=.535\textwidth]{images/den_small_RdBu_R0_I2_half_15.png}
%    \includegraphics[width=.535\textwidth]{images/den_small_RdBu_R0_I2_half_17.png}
%    \includegraphics[width=.535\textwidth]{images/den_small_RdBu_R0_I2_half_20.png}
%    \caption{Close-up  to the star. Although the envelope changes dramatically between snapshots, the central component stays steady through out the simulation. Here $t_{end}=14 \ \text{years}$.{\bf CM n instead of $\rho$}.}
%    \label{fig:close up star}
%\end{figure}

\begin{figure}[ht]
    \centering
    %\subfloat[]{\includegraphics[width=.52\textwidth]{images/ohneGridDen.png}\label{fig:sub1}}\hskip1ex
    %\subfloat[]{\includegraphics[width=.515\textwidth]{images/ohneGridRadius.png}\label{fig:sub2}}
    \includegraphics[width=.52\textwidth]{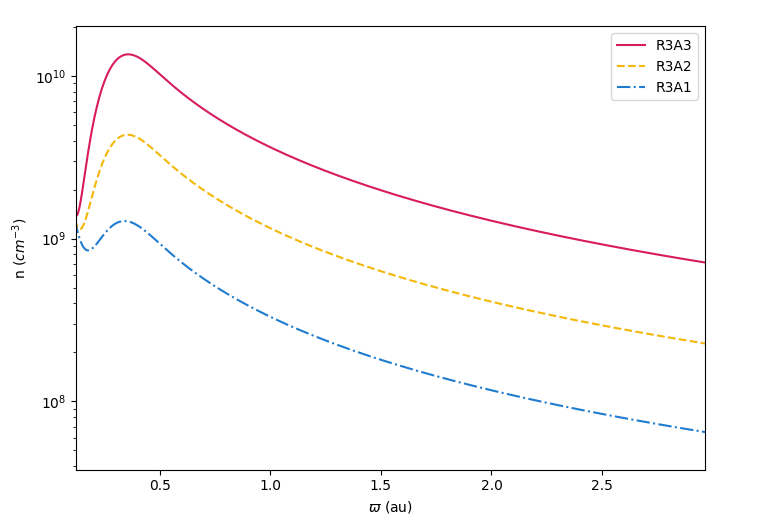}
    \includegraphics[width=.515\textwidth]{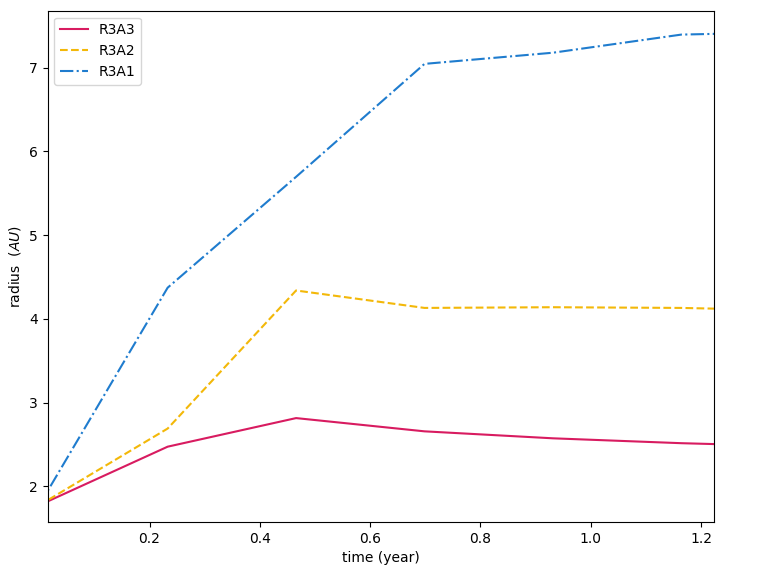}
    \caption{(a) The effect of different normalizations on the disk density. (b) Evolution of the radius with respect to time. Lowest values of the radius are from R3A3, the highest jet radii are given by R3A1. And in between the jet radii from R3A2. the radii are extracted at the radial boundary, close to  $10 \ \rm{au}$ on the axis.}
    \label{fig:density at surface}
\end{figure}

In Fig.\ref{fig:density at surface}, we plot the jet radii as function of time. We find that the simulations reach their final jet radius only after a few stellar rotations, between $8$ and $9$ stellar revolutions for all simulations. The number of stellar rotations needed to achieve jet stability depends on the size of the computational box. From Fig.\ref{fig:density at surface}, we see that the lowest density normalization almost matches the slope of the stellar solution at the equator. In this case, looking at the forces profiles near the star, we see that the pressure gradient takes over the Lorentz force. Conversely, for the smallest jet, the Lorentz force is dominant. Doing the same analysis on the forces at a higher $z$ shows an overall agreement between the three cases. This implies that the steady state reached mainly depends on the force balance at the boundary condition on the equator. The reason we notice a predominant pressure gradient as we decrease density is due to the fact that by decreasing density the magnetic field strength also decreases.

\begin{figure*}[ht]
    \centering
    \includegraphics[width=.41\textwidth]{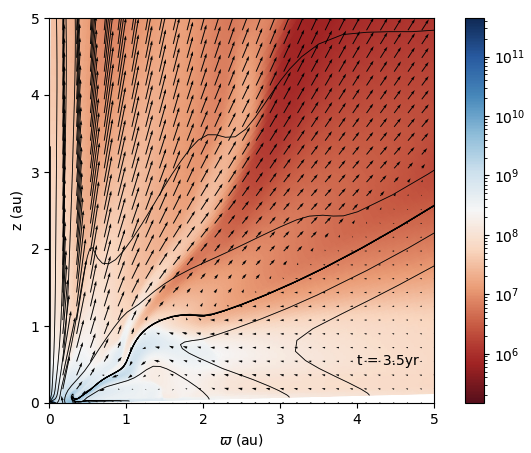}
    \includegraphics[width=.41\textwidth]{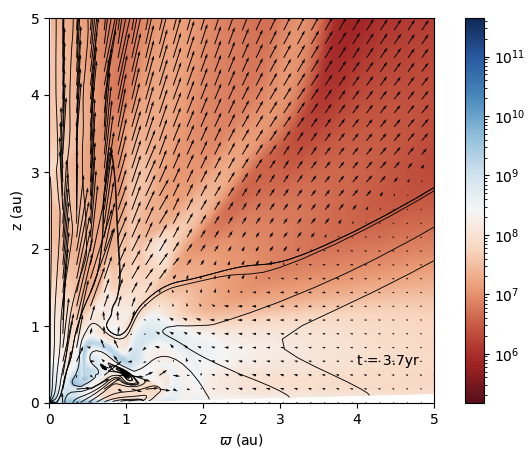}
    \includegraphics[width=.41\textwidth]{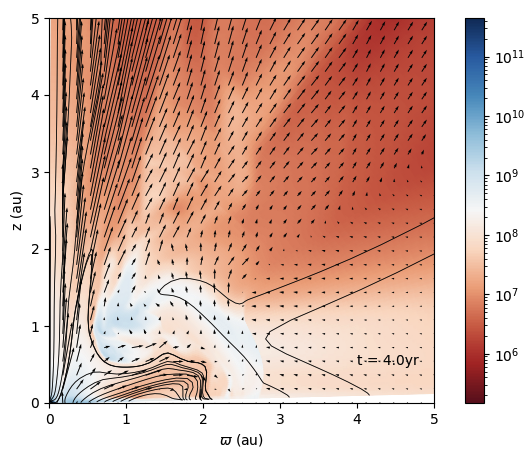}
    \includegraphics[width=.41\textwidth]{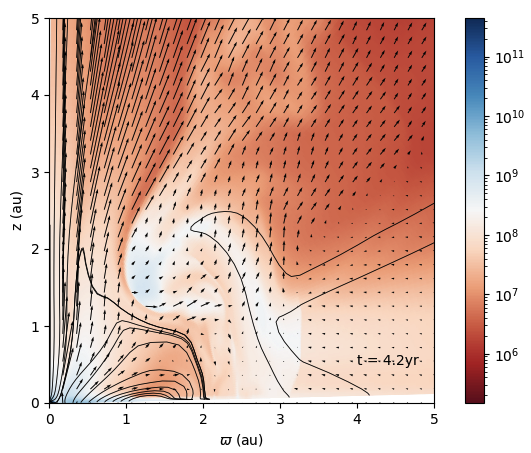}
    \includegraphics[width=.41\textwidth]{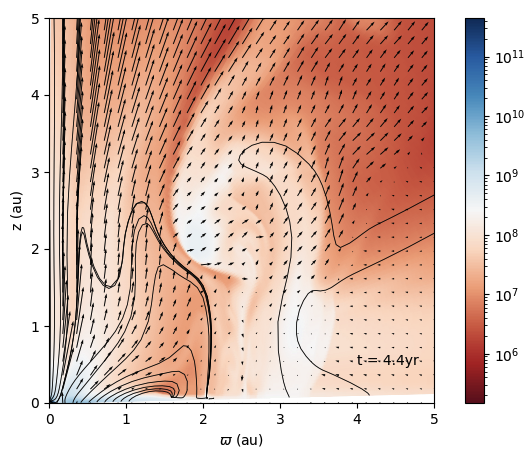}
    \includegraphics[width=.41\textwidth]{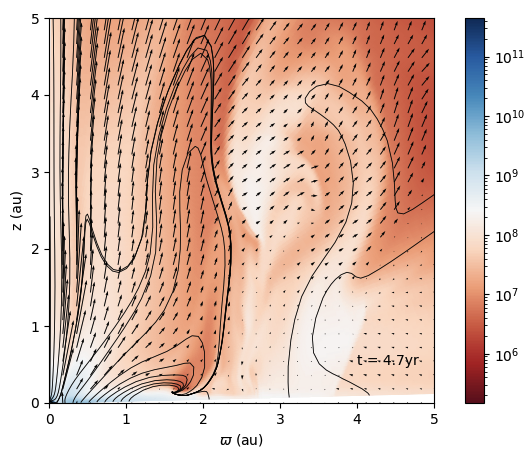}
    \caption{Appearance and evolution of a magnetospheric ejection (hereafter ME). The ME creation and ejection processes span over a year. At t=0, we have the initial state. In all plots, the black lines trace the magnetic field, the black arrows trace the velocity field, and the color map traces the density. At $t=4~\rm years$, we see the formation of two shocks on the axis, which propagates along the jet. For this simulation, $R_c = 0$ and $I_h = 10^{-2}$.}
    \label{fig:ME production}
\end{figure*}
%\begin{figure*}[ht]
%    \centering
%    \includegraphics[width=.45\textwidth]{images/den_small_RdBu_R0_I2_half_15.png}
%    \includegraphics[width=.45\textwidth]{images/den_small_RdBu_R0_I2_half_16.png}
%    \includegraphics[width=.45\textwidth]{images/den_small_RdBu_R0_I2_half_17.png}
%    \includegraphics[width=.45\textwidth]{images/den_small_RdBu_R0_I2_half_18.png}
%    \includegraphics[width=.45\textwidth]{images/den_small_RdBu_R0_I2_half_19.png}
%    \includegraphics[width=.45\textwidth]{images/den_small_RdBu_R0_I2_half_20.png}
%    \caption{Close-up  to the star. Although the envelope changes dramatically between snapshots, the central component stays steady through out the simulation. Here $t_{end}=14 \ \text{years}$.}
%    \label{fig:close up star}
%\end{figure*}

We observe that as we increase density, the heating tank available for a volume is smaller. It effectively reduces the energy delivered to the plasma. As a result, 
outflow regions narrow and plasma material returns to the disk. This bends the magnetic field lines above the disk and increases the magnetic tension. In this configuration, we see the formation of periodic magnetospheric ejections (MEs) happen whenever the magnetic tension is high. The opening and closing of the field lines expel matter outwards with a timescale of around $7 \ \text{years}$. These episodic outflows cross the fast magnetosonic surface, which makes them decoupled from the base of the outflow. In short, the observed MEs are not entirely ballistic outflows and constitute a part of the low velocity component of the jet. Fig.\ref{fig:ME production} shows the kind of episodic outflows in our simulations, this case corresponds to $R_c = 0$ and $I_h = 10^{-2}$ (R0\_I2). Consequently, after MEs are launched, they travel and affect the jet itself. As magnetic density grows near the polar axis the jet collimates more. Then, when matter is released, a wobbling structure
can be seen along the jet in accordance with observations \citep{lopez2003proper,pyo2003adaptive,takami2020possible}. Observed knots coming from young stars jets are attributed to mass and velocity variations during ejection \citep{purser2018constraining}, which is the case here. This configuration does not reach a steady state, and has more abrupt periodic ejections than what can be seen in \cite{ZanniFerreira13}, and \cite{ireland2022effect}. Like in \cite{ZanniFerreira13}, MEs mostly include disk material and only a very small percentage of stellar material mixes in the MEs in a stable non-irruptive way. The more energetic part of the ME spans over years and is constituted of disk material only. If compared to \cite{ireland2022effect}, models where strong MEs are presents are accompanied by a pause in accretion. In our case, even with strong MEs, material keeps accreting. Without fail, accretion and outflow close to the star stay close to the initial condition and matches the final steady state reached in \cite{sauty2022nonradial} for $\text{case} \ C$. We discuss the effect of increasing accretion in Sec.\ref{sec:increasing accretion}.

Since the magnetic field lines at the disk protect the central engine, the accretion rate and pressure balance are not modified. Thus, the close up is very similar to those found in \cite{sauty2022nonradial}. We can notice a small difference due to a denser and stronger magnetic field at the star vicinity. Compared to simulations from \cite{sauty2022nonradial}, the stellar jet is more collimated. Fig.\ref{fig:close up star}, contains a close up of the inner region with the same color scale and box size.

\subsection{Effect of heating}
To gauge what effect heating has on the jets dynamics, we can look at the radius, force balance, electric currents, and integrals of motion. Still  we choose to show physical tracers that can help the set of parameters that better fit observations to be determined. We will try to address the following:
\begin{itemize}
    \item The evolution of the jet radius with heating.
    \item The size of the jet emitting disk.
    \item The mass-loss evolution with the heating.
    \item The range of temperatures obtained.
\end{itemize}

\subsubsection{Radius evolution with heating}
\begin{figure}[ht]
    \centering
    \includegraphics[width=.485\textwidth]{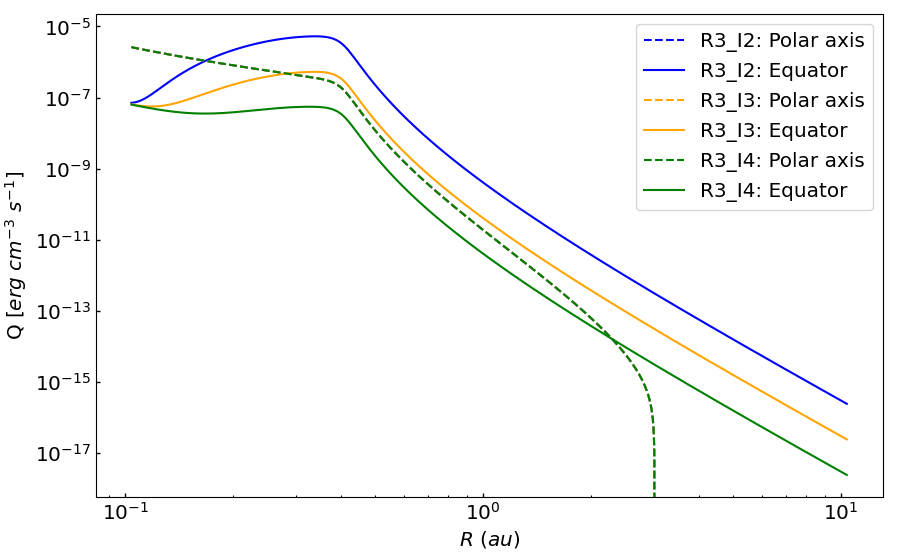}
    \includegraphics[width=.5\textwidth]{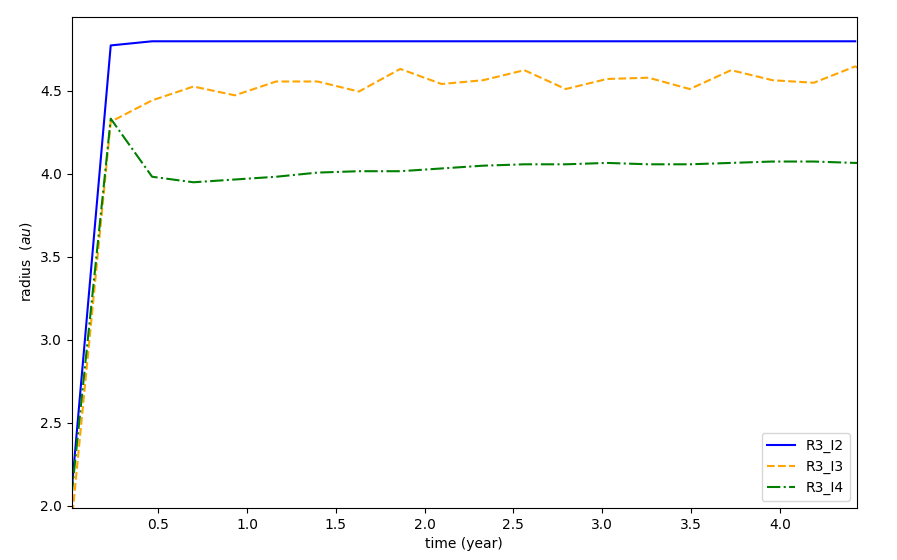}
    \caption{Heating rates (upper plot) and the corresponding jet radius (lower plot) extracted from the simulations for a fixed $R_c$ and a variable $I_h$. For all the cases, the stellar component is unchanged and matches the green dotted curve. In the upper plot, the dotted lines express the heating extracted at $\theta \sim 0$, and the solid lines are for values of the heating on the equator ($\theta \sim \pi/2$).}
    \label{fig:heating v radius}
\end{figure}

Referring to Fig.\ref{fig:param space}, we have two ways of controlling the jet width. We can either fix $R_c$ and vary $I_h$, or conversely fix $I_h$ and vary $R_c$. We first look at how the radius evolves in Fig.\ref{fig:heating v radius} where $R_c$ is fixed. We put the cut off of the analytical heating  at $0.3 \ \rm{au}$ and vary $I_h$ (in Tab.\ref{tab:simID} this concerns all the simulations starting with R3\_\#). We follow the spine evolution through the passive tracer introduced previously. The larger and slower disk jet is hard to probe with the tracers. The reason is that the launching region at the base varies from one solution to another. The launching region of the jet is not the same for all cases, and depends heavily on the heating intensity. If we want to understand the evolution of the jet and disk wind ,we need to use the density maps directly. These maps show the gap between the disk and the jet. They show as well the standing collimation shocks that appear inside the computational domain for some of the parameters. Fig.\ref{fig:heating v radius} shows the values the heating rate takes and, consequently, the jet radii. The stellar heating is common to the three cases, and only the wind heating varies by changing $I_h$. We note that as we increase the wind heating, the jet radius also increases. We can explain such a behavior by the additional radial pressure introduced by the heating process. This pressure pushes the disk wind at a higher angle out of the disk, which results in a less collimated jet.

\begin{figure}[ht]
    \centering
    \includegraphics[width=.485\textwidth]{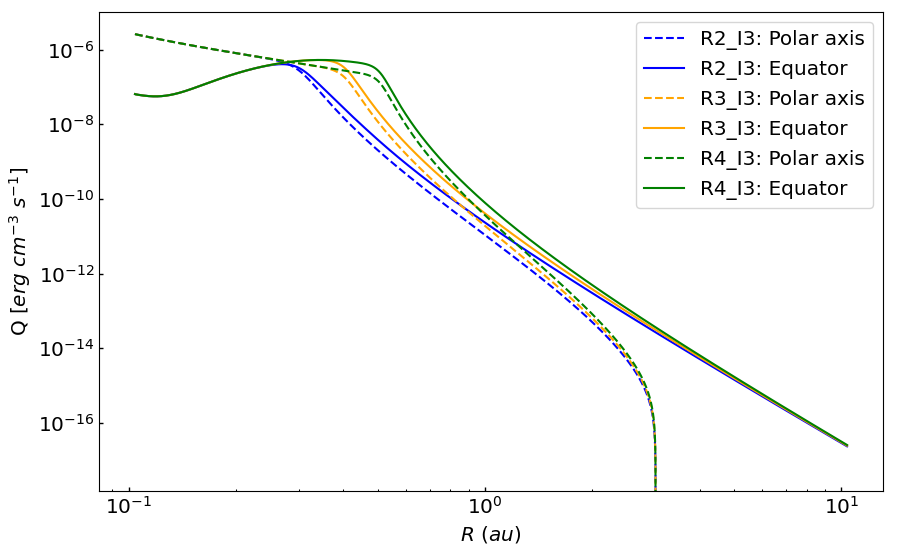}
    \includegraphics[width=.49\textwidth]{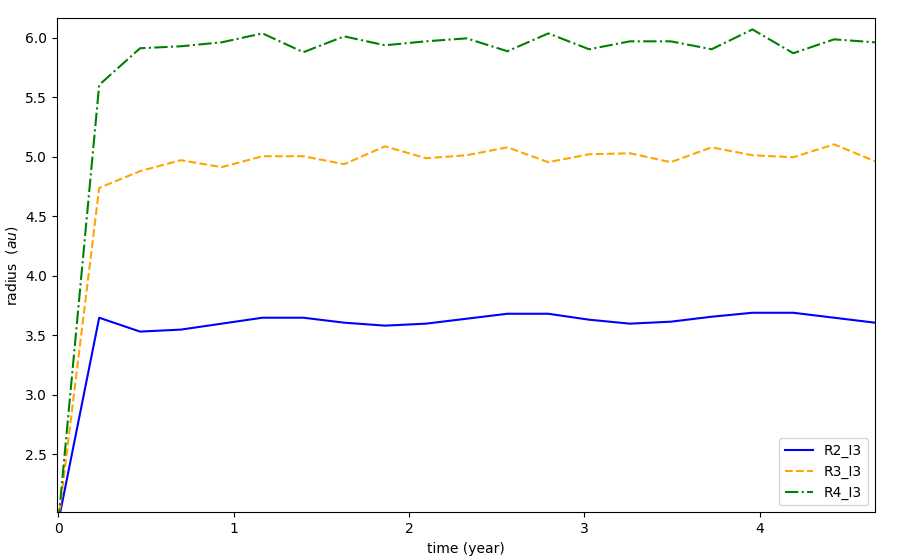}
    \caption{Same plot as in Fig. \ref{fig:heating v radius} but for a fixed $I_h$ and a variable $R_c$. }
    \label{fig:heating v radius2}
\end{figure}

In Fig.\ref{fig:heating v radius2}, we see that by selecting a fixed value for $I_h$ and varying $R_c$, the wind also acquires extra heating. This is expected as we extend the heating farther away from the star for both components. If we only increase the heating of the stellar component, this leads to a lower increase in the jet radius. It means that only varying $R_c$ does not isolate the contribution of the stellar heating. When we isolate it we, once again, see that the farther we heat up the stellar atmosphere, the larger the jet radius is. It is in concordance with our conclusions so far. 

Same as for the spine jet, the disk jet opens up as more heating is added. Its opening angle is larger by definition. For the wider disk jet, it is difficult to define a formal characterization of its radius. For that reason, we rely mainly on the spine component to extract information on the radius. We nonetheless notice a number of generic behaviors common to all cases. The disk jet has the same topology as the spine jet no matter the heating, for example, if the spine is paraboloidal, the disk jet is as well. This holds for the two other cases. By extracting the mean heating rate from $R=0.085 \ \rm{au}$ (outside the accretion region) to $R= 1 \ \rm{au}$ and above the disk up to $\theta = 1.3 \ \rm{rad}$, we see that above $Q = 3 \cdot 10^{-6} \ \rm{\ erg \ cm^{-3} \ s^{-1}}$ the low density corridors separating jet and disk are more pronounced. The outflow separation region is lower than $10^{-5} \ \rm{cm^{-3}}$ in density. The lower the heating, the narrower and denser the medium.

\begin{figure}[ht]
    \centering
    \includegraphics[width=.49\textwidth]{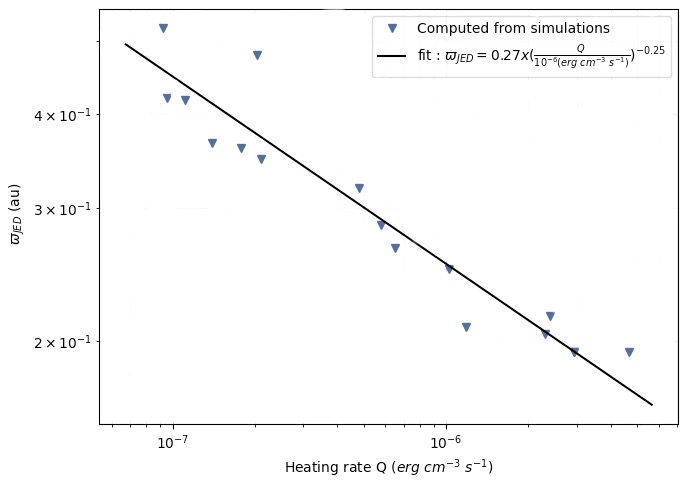}
    \caption{Fitting the disk jet radius at the base of the launching. All cases with surface accretion were excluded.
% NV:  try to explain this simple power law...
    }
    \label{fig:radius fit}
\end{figure}

Overall, we can fit the evolution of the disk jet radius emitted at the base using a power-law of the average heating rate. We exclude in Fig.\ref{fig:radius fit} all the cases that have a surface accretion as they do not reach a steady state. To fit the data we keep 18 simulations out of 26. All nonsteady jets are found on the upper left corner of the parameter space (Fig.\ref{fig:param space}). We argue it is due to the weaker contribution of the stellar heating. Even if they overlap with stationary cases when looking at the average heating rate. Having a stronger stellar heating component expels more matter at initial steps, which in turn lightens the wind in terms of mass. If enough material is expelled in the launching region it becomes easier for the centrifugal and Lorentz forces to keep plasma from migrating closer to the disk. It is worthwhile noting that the values of the average heating rate depend heavily on the area of the grid we extract the information from. As can be seen from the upper plots of Fig.(\ref{fig:heating v radius},\ref{fig:heating v radius2}), most of the heating is close to the star. 
The value of heating presented in these plots is sensitive to the size of the region over which we take the heating average. 
This being said, even if the value of the heating rate changes the scaling law does not, following $\varpi_{JED} \propto Q^{-0.25}$. This power law of $-1/4$ can be explained by looking at the thermal pressure added by the heating. The plasma is accelerated more. Given the disk is a boundary that refills the wind at a constant rate, the region, which has enough matter to form a disk jet, also decreases in size. 
%As the disk radius, the mass loss rate increases like the logarithm of the disk radius (following the self similar law). The heating rate on the equatorial plane decreases with distance in a complexe way. However, in average, it roughly decreases with $\varpi^{-4}$ on Figs.\ref{fig:heating v radius},\ref{fig:heating v radius2}. This may explain the fits we obtain.

%% The power laws are quite complicated Q in average varies like varpi^-4 see Fig 10 as it decreases as varpi^-6 Mdot varies as ln varpi so Mdot varies like 1-ln varpi so finally it is probably better to not say anything

We assume that no matter how small the heating, the flow does not fall back onto the disk but goes out. Under this assumtion, we can infer from Fig.\ref{fig:radius fit} that below $2.56 \cdot 10^{-8} \ \rm{ erg \ cm^{-3} \ s^{-1}}$ the radius of the jet emitted by the disk is larger than $0.7$ au. For a jet to form, MHD requires the gas to support an electrical current, which in turn needs charge carriers. We need processes that can sufficiently ionize the gas outwards. Typically, a Class II star can thermally ionize up to $ \sim 0.1$ au, above the disk, which only represents a small fraction of the disk. 
In accreting systems, the magneto-rotational instability (MRI) also needs ionization further outward in the disk. Otherwise a dead zone appears in the mid-plane of the disk lowering accretion. The degree of ionization is uncertain and varies with the strength of each ionization process. For RY Tau, we assume the ionized part of the disk that can produce a jet between $0.1-0.7$ au. This makes $2.56  \cdot  10^{-8} \ \rm{erg \ cm^{-3} \ s^{-1}}$ a lower limit for heating.

The ionization level needed to initiate MRI and to create a medium with enough charge carriers to facilitate ejection is easily modeled up to $\sim 0.3 \ \rm{au}$ through energetic particles produced during reconnection events of the stellar magnetic field with the disk. 
%\textcolor{magenta}{CS: But are jets emitted beyond this value? 0.5 au i already enough?} 
\cite{brunn2023ionization} estimate the ionization generated by protons, electrons, and secondary electrons accelerated from solar flares. They find ionization rates 6 times higher than what is proposed by other sources such as stellar X-rays, and  galactic cosmic rays \citep{rab2017stellar}. This extra energy contributed by flaring is sufficient to maintain MRI on secular scales, and generates heating rates that match what we have parameterized in our model \citep{Brunn2024}. Above $0.3 \ \rm{au}$ we need other injection sources and/or sources of ionization.

\subsubsection{Disk wind mass loss}
\begin{figure}[ht]
    \centering
    \includegraphics[width=.48\textwidth]{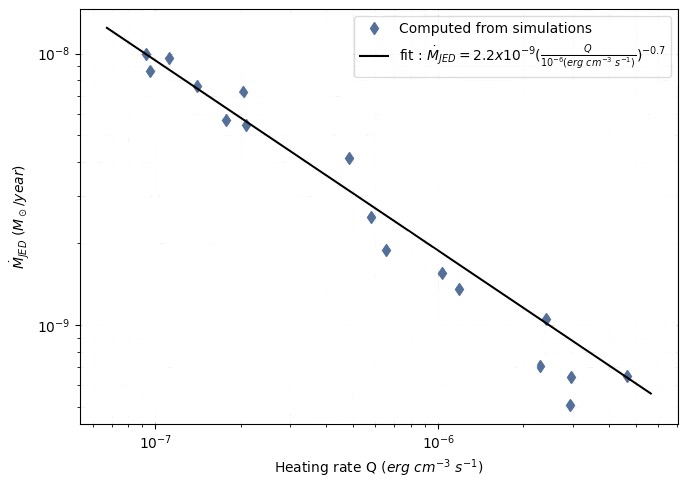}
    \caption{Plot of the mass loss with the heating rate. The mass loss as function of the heating can be fitted by a power law.
    }
    \label{fig:masslossVsheating fit}
\end{figure}

The mass loss rates follow the same trend as the radius as the heating increases (Fig.\ref{fig:masslossVsheating fit}). This is expected, since by reducing the launching region, the jet mass loss rate should also decrease. We can extend the analysis  on the upper limit of the disk ejection radius. We can infer an upper boundary for the mass loss rate as well. We obtain that the disk jet should not expel more than $2.8 \cdot 10^{-8} \Msunyr$. We infer that result from Fig.\ref{fig:masslossVsheating fit}. 
The mass loss scales as $\dot{M}_{JED} \propto Q^{-0.7}$. \cite{anderson2005structure} investigate how the wind structure is influenced by the mass loading. They measure the cylindrical radius (called in their paper, $\varpi_j$) along the field containing $\% 50$ of the mass flux, which originates from the same disk region in all their simulations. The increase in collimation is found to be slow, and they fit $\varpi_j$ using a power-law in $\dot{M}_{\varpi_j}$ bearing a rough relation that goes as follows $\varpi_j \propto \dot{M}^{0.1}_{\varpi_j}$. We obtain by fitting $\varpi_{JED}$ and $\dot{M}_{JED}$, the following rule, $\varpi_{JED} \propto \dot{M}_{JED}^{0.33} \propto (\dot{M}_{tot} - \dot{M}_{spine})^{0.33}$. Contrarily to \cite{anderson2005structure}, we do not fix the ejection radius to where the magnetic field line accounts for half the mass flux, but we instead measure the whole mass flux coming from the region of the disk that produces the jet.

\subsubsection{Magnetospheric outflows}
Magnetospheric
%\textcolor{magenta}{CS the "X-type" terminology is really controversial. If it is due to the zone between the disk and the magnetosphere, it can be called magnetospheric mass ejection} 
winds and jets expelled by a larger portion of the disk are both observed in the parameter space. They depend on the amount of heating introduced. To obtain jets with launching regions above $0.2 \ \rm{au}$, we need heating rates lower than $\sim 3 \cdot 10^{-6} \ \rm{erg \ cm^{-3} \ s^{-1}}$. Cases with launching regions below $\sim 0.2 au$ have a morphology similar to that of X-type winds modeled by \cite{shuetal94}. 

An important result from this study is that the more the heating we input in the disk wind, the wider the jet is. If too much heating is implemented, the outflows are too wide to be called jets. This is the case for the quasi-conical solutions. The outflows become radial in the form of a wind. The eventual
%CS: eventual ne veut pas dire éventuel, mais "qui s'en suit" 
velocities from the disk outflows are too high ($\sim 1000 \rm{km \ s^{-1}}$).

\begin{figure}[ht]
    \centering
    \includegraphics[width=.41\textwidth]{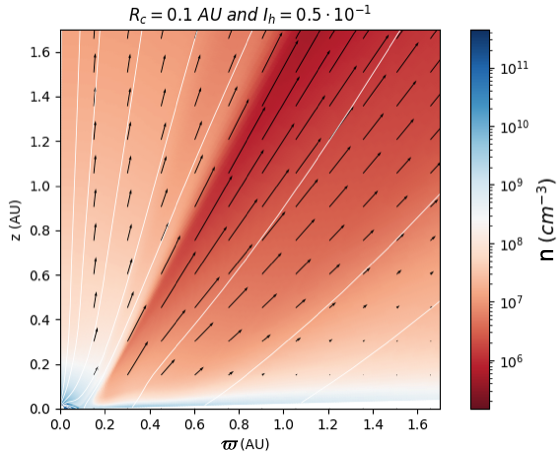}
    \includegraphics[width=.41\textwidth]{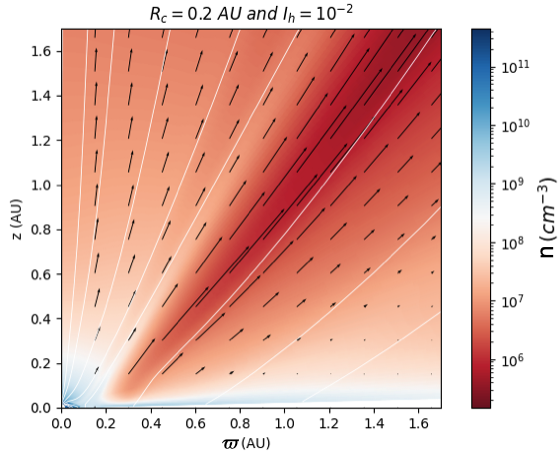}
    \includegraphics[width=.41\textwidth]{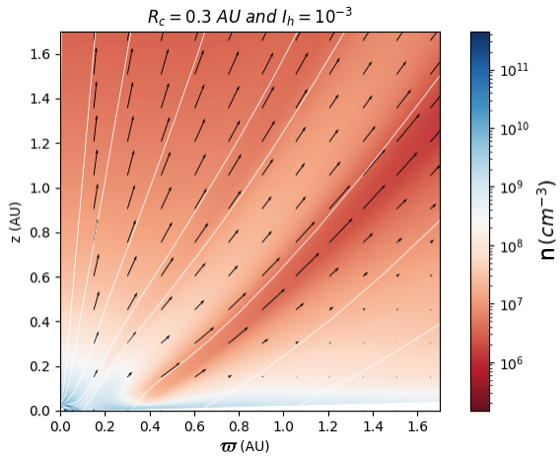}
    \caption{Width of the conical winds given the chosen parameters. The simulations are steady.}
    \label{fig:intermediateRho}
\end{figure}

The X-type winds observed in this work emanate from the corotation radius ($r_c$). The funnel stream starts at a truncation radius ($r_t$) close to $r_c$ ($r_t \approx r_c$). This means the system is in a configuration that does not add angular momentum to the star through the closed magnetic lines in the accretion zone. Since $r_c$ and $r_t$ are imposed by the boundary conditions, the angular momentum is not affected by the changes that may occur in the disk. The exception would be if enough matter migrates closer to the star and interferes with accretion. However, we do not get such configurations, because they lead to a steeper time increment and make these scenarios difficult to obtain and study. Similar to \cite{romanovaetal09} we get conical winds with wide opening angles. Unlike their work, we have a fast inner jet component on the axis. Additionally, the outflows are steady with no inflation of the magnetic field lines. Although, the extraction region of these conical outflows are constrained to a small spatial range, the mass loss rate is on the order of $ \sim 10^{-9}$. As a comparison, the mass loss rate of the stellar engine is $3.0 \cdot 10^{-9} \Msunyr$. In the upper plot of Fig.\ref{fig:intermediateRho}, the mass loss from the conical wind only is $1.3 \cdot 10^{-9} \Msunyr$. The middle plot produces $2.3 \cdot 10^{-9} \Msunyr$. The last one gives $5.5 \cdot 10^{-9} \Msunyr$. These values were obtained by computing the mass loss rate between two colatitudes that contain the conical component, namely,
\begin{equation}
    \dot{M} = 2 \pi R^2 \int_{\theta 1}^{\theta 2} \rho V_R \sin \theta d \theta.
\end{equation}

\noindent We hence determine what constitutes the global mass loss of the disk jet by looking at the mass flux passing through two density minimums.

These solutions have a shortcoming due to the regions below $10^{-5} \ cm^{-3}$ that separate the jet and the disk wind. They appear because the heating rate is strong enough to expel matter faster than the disk can repopulate the region. The resulting temperature is on the order of one million degrees. Density decreases faster than pressure, which leads to these high temperatures. The temperature we derive from the simulations are effective values that contain extra terms that do not describe the gas pressure only. Thus, because the pressure has extra terms, we argue that our temperatures are overestimated when density is below $10^{-5} \ \rm{cm^{-3}}$. 
%\textcolor{magenta}{CS: what do you mean by the last sentence?}

\subsection{Electric circuit}
The poloidal electric current responsible for the collimation leaves the disk to enter again at closer radii. The current direction has to change in the jet as a whole. The current density $J_p$ is outwards in the outer disk and inwards in the jet. Anchored field lines have, consequently, two regions, one where the field lines close on the axis, the other where the magnetic field lines decollimate. Since the initial bow shock leaves the simulation domain, the global electric circuit is maintained by the boundary conditions. 

Near the axis $ \varpi B_\phi$ is necessarily zero. Its value decreases as we get farther from the axis, and reaches a minimum at the end of the jet. This pattern is similar for all the cases we studied. We begin to see differences outside the jet. For winds with the highest heating rates, very low density regions appear, which disconnect the disk electric circuit from the jet. $\varpi B_\phi$ profile increases after leaving the jet to reach approximately zero in all the low density region. 

The paraboloidal and conical jets present the expected butterfly shape. The overall circuit is well behaved, and without any particular feature that would help identify the morphology of the jet. \cite{heyvaerts1989collimation} focused on the electric poloidal current present at infinity. Their results indicated that any stationary axisymmetric magnetized jet will eventually converge into either paraboloids or cylinders, asymptotically from the source. The outcome depends on whether the asymptotic electric current vanishes or takes finite values. This fundamental theorem was later extended by \cite{heyvaerts2003global} to incorporate considerations of current closure and its influence on the solution geometry. The actual amount of current left is not known since linking current value to the source is not simple to make in numerical works.

\subsection{Increasing accretion rate}\label{sec:increasing accretion}
\begin{figure}[ht]
    \centering
    \includegraphics[width=.45\textwidth]{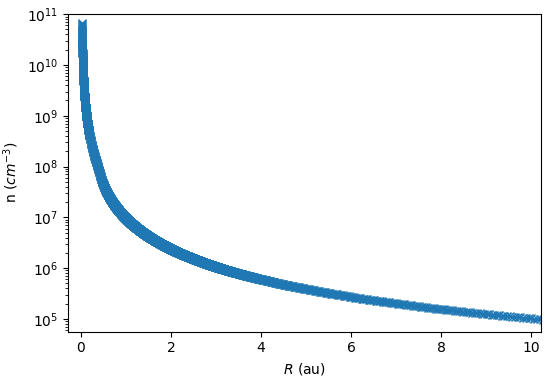}
    \includegraphics[width=.45\textwidth]{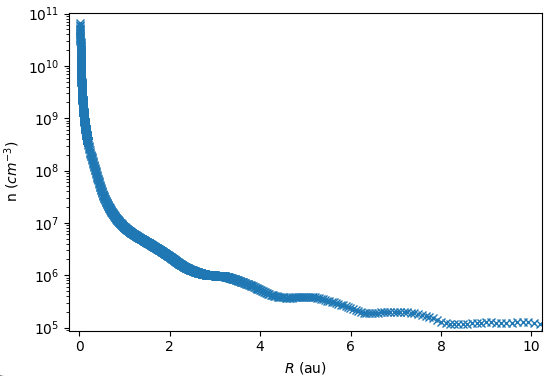}
    \caption{Number density of the R3\_I4 with an accretion rate of $1.3 \cdot 10^{-8}$ in the upper plot and $3.1 \cdot 10^{-8}$ in the lower one. We fix $\theta$ at $\sim 0$ to follow the evolution of density along the polar axis.}
    \label{fig:effect of a higher accretion rate}
\end{figure}

Two accretion rates were chosen to quantify the effect of changing accretion rate on the global jet. Observed accretion rates of intermediate mass young stars ($1.5-3 \ M_\odot$) are between $10^{-8} - 10^{-7} \Msunyr$ \citep{garufi2019sphere}. All the simulations shown so far have an accretion rate of $1.3 \cdot 10^{-8} \ \Msunyr$. Increasing the accretion rate by a factor of three creates a wave pattern on the spine jet (Fig.\ref{fig:effect of a higher accretion rate}). This wavy stellar jet evolution is the consequence of magnetospheric ejections. Similarly to cases D, D.1, and D.2 from \cite{sauty2022nonradial}, increasing the mass accretion rate in the magnetospheric columns introduces quasi-periodic stable magnetospheric ejections that remove angular momentum from the star and the disk. It would prove highly challenging to differentiate observationally between the two accretion rates due to the substantially greater uncertainties in the measurements compared to the variation identified in our simulations. Nonetheless, they serve as proof for magnetospheric ejections in the case of an increased accretion rate, where up to $25 \%$ of accretion mass loss feeds the spine outflow. 

\cite{grinin1992young} proposed that RY Tau exhibits irregular eclipse-like dips in brightness, attributed to fluctuations in obscuration resulting from circumstellar dust. In the same spirit, \cite{petrovetal19} suggest a bimodal behavior of RY Tau. One where the brightness of the star increases, and so does the variability of ejection. The other mode is called quiescent, and is characterized by a fainter source. the authors explain the change in luminosity by a changing dust structure around the star. With more activity, the geometry of the dust changes and clears the star line of sight.

We suggest that in addition to the steady variability introduced by increasing accretion onto the star \citep{sauty2022nonradial}, an increase in matter infall farther in the disk also contributes. This episode in the early life of the star would let dust migrate closer to the star. Then with the eruptive ejection that occurs after, dust is evacuated. In the simulations where we see this scenario, the period for the cycle to complete is between 1 and 2 years (Fig.\ref{fig:ME production}), followed then by a calmer phase, where plasma that has not been evacuated in the jet fall back and repopulates the disk farther outward.     

\subsection{Standing recollimation}\label{sec:standing recollimation}

\begin{figure}[ht]
    \centering
    \includegraphics[width=0.5\textwidth]{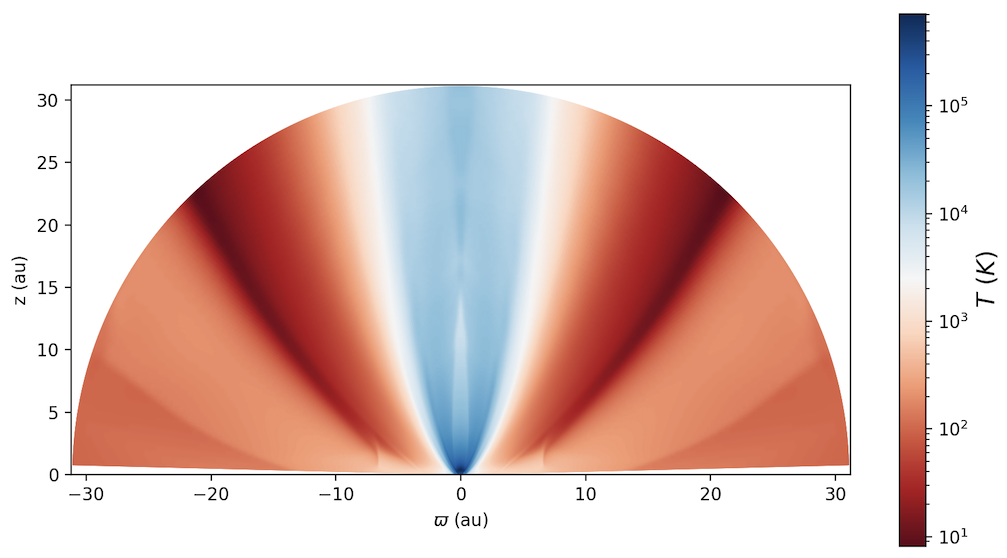}
    \caption{Temperature plot of R3\_I4. We extended the computational grid for this simulation to $30 \ \rm{au}$ to capture the standing shock. the maximum temperature is around $10^6 \ K$ and can be found below $1 \ \rm{au}$, near the star.}
    \label{fig:temp shock}
\end{figure}

\begin{figure}[ht]
    \centering
    \includegraphics[width=.48\textwidth]{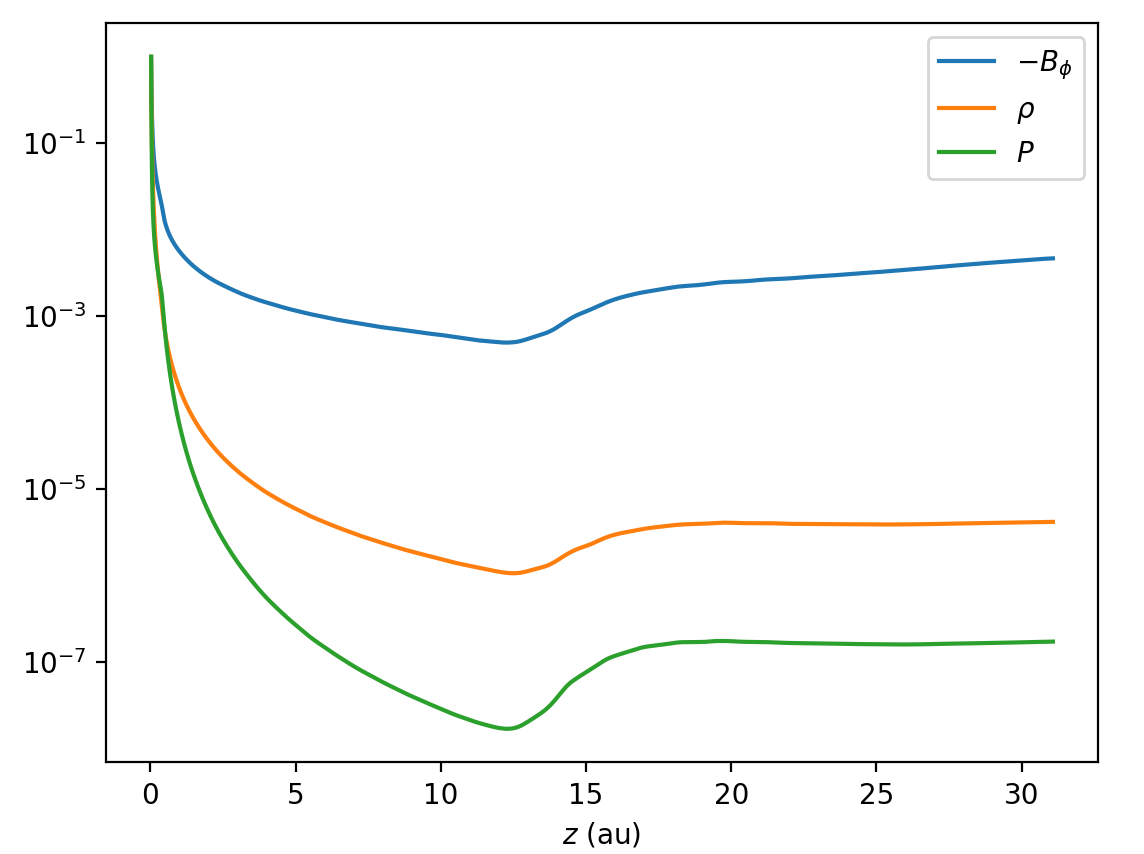}
    \caption{Plot of the toroidal magnetic field, density, and pressure, along the jet axis. All quantities are normalized such as to fit in the same plot.}
    \label{fig:jump}
\end{figure}

We note the presence of a standing shock structure in R3\_I4. This happens when the flow hits the axis at a velocity greater than the fast magnetosonic velocity, resulting in a jump in the flow quantities. After the shock, the magnetic fields are refracted outwards. We can better see the collimation and appearance of a shock in the temperature plots (Fig.\ref{fig:temp shock}). The outflow is collimated through inwards Lorentz force that become larger as the jet widens. The flow reaches the axis at $R \sim 15 \ \rm{au}$, resulting in a shock surface that has the shape of a cone along the polar axis. In Fig.\ref{fig:jump} we show that a jump in the flow occurs after at the shock. The compression ratio $\rchi$, defined as the density post- and pre-shock, is around $2$. Given the uncertainty on when the shock front starts, we obtain a compression factor in the interval $[2,3]$.The maximum compression rate possible for an adiabatic hydrodynamical 1D flow is $\rchi_{0} = \Gamma + 1 / \Gamma - 1 = 4$. 

As can be seen in Fig.\ref{fig:temp shock}, the shock is not normal to the flow. In this case, the shock is an oblique one. To determine its properties we have to look at the normal components to the shock surface.   

\begin{figure}
    \centering
    \includegraphics[width=.48\textwidth]{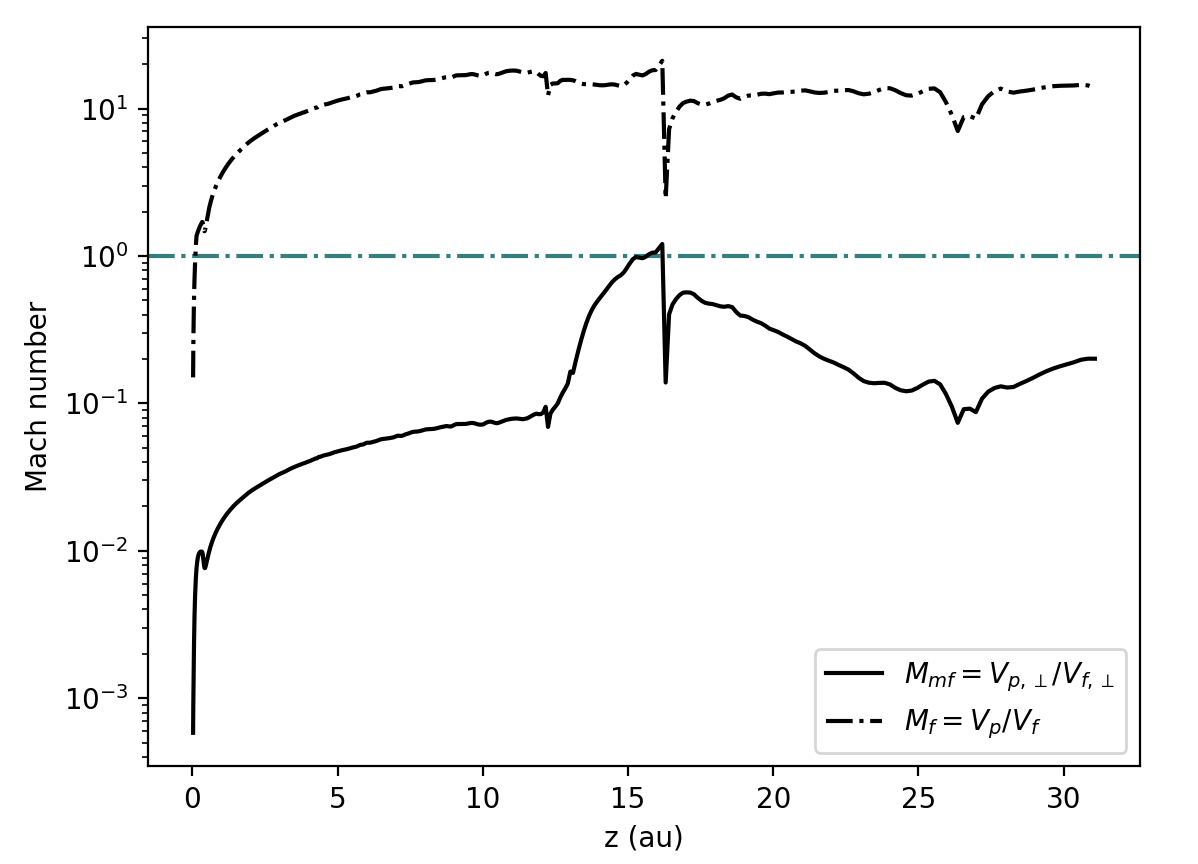}
    \caption{Plot of the perpendicular fast magnetosonic Mach number $M_{mf}$ (same terminology as VTST00) and with $M_f$ describing the Mach number parallel to the flow. The blue dash-dotted line corresponds to the fast magnetosonic limiting characteristic where the perpendicular fast Mach number equals to $1$. The $M_{mf}$ and $M_f$ shown are computed along $\theta \sim 0.3$.}
    \label{fig:jump Mach fastPerp}
\end{figure}

The outflow is supersonic and super-Alfv\'enic, and the sound speed is computed in the jet using $C_s = \sqrt{\partial P/\partial \rho}$, along the polar axis. The toroidal component of the field is the dominant one ($B_\phi / B_p >> 1$). To best capture the nature of the shock, we compute the fast magnetosonic Mach number using the velocity perpendicular to the shock  $M_{mf} = V_{p,\perp} / V_{fast,\perp}$. We call it the modified fast Mach number. If, on the other hand, we look at the poloidal fast magnetosonic Mach number, $M_{f} = V_{p} / V_{fast}$, we see that the flow remains super-fast even where the shock lays. The two quantities are computed along $\theta \sim 0.3 \ \text{deg}$ as there is no perpendicular velocities on the axis by construction. Fig.\ref{fig:jump Mach fastPerp} shows a clear jump of the velocity from the super fast regime before the shock to a sub-fast velocity after. 
%Similarly, \cite{Matsakosetal2008} found for disk wind self-similar models that the shock happens soon after $V_{p,\perp} / V_{fast,\perp} = 1$ and not $V_{p} / V_{fast} = 1$. 
The shock wave angle ($i$) is difficult to determine precisely given the grid resolution. We find $i$ between 0.5 and 2.6 degrees. Consequently, the shock occurs when the $M_{mf}$ is between $1 - 1.7$. In Fig. \ref{fig:jump Mach fastPerp}, we plot the modified fast Mach number, using the perpendicular velocities, for a cone of $0.5\deg$.

Using an angle of $0.5\deg$, we find that, at the shock location, the modified fast Mach number is equal to one. This coincides with the results of \cite{Matsakosetal2008} where the shock converges to the fast critical surface. The fact that the poloidal fast Mach number remains above one, shows that the real critical surface is on the modified fast critical surface, which was demonstrated by \cite{Bogovalov94} as a singularity. Yet we extend here the conclusion of \cite{tsinganos1996critical} for axi-symmetric and self-similar solutions. The real critical surface is only the modified critical surface. For the critical surfaces, the component of the flow velocity perpendicular to the directions of axi-symmetry and self-similarity equals the slow/fast MHD wave speed in that direction. The modified fast surface coincides with the real MHD horizon, while the poloidal fast surface is the MHD ergosphere, analogously to black holes, see \cite{Carter68}.

Recollimated plasma is systematically seen in R3\_I\# (all simulations starting with R3). The characteristic flow toward the axis happens inside the computational grid, but the shock itself can occur outside. For the closest shocks, they happen between $5$ and $15 \ \rm{au}$. By lowering the heating rate in the disk wind, the shock appear closer to the star.

%\subsection{Fitting the heating on the disk surface}
%\subsection{heating vs. mass-loss rates}
%\begin{figure}[ht]
%    \centering
%    \includegraphics[height=.3\textheight]{images/heatrate_massloss.png}
%    {\footnotesize (a)massflux on a horizontal surface with Vz}
%    %\includegraphics[height=.3\textheight]{images/massloss_Vp.png}
%    %{\footnotesize (b)massflux on a horizontal surface with Vp \textbf{y axis is weird}}
%    \includegraphics[height=.3\textheight]{images/angle_VzVp.png}
%    {\footnotesize (c)Angle between Vz and Vp}
%    \caption{Plots made by taking the mean values between $0.085 \ AU$ to $1 \ AU$.}
%    \label{fig:QvsM}
%\end{figure}

\section{Discussion}\label{sec:discussion}
\subsection{Comparison to other simulations}
We compare our work to other platform and global 2.5D simulations of nonrelativistic jets. Three-dimensional simulations would allow us to explore a larger range of possible instabilities and phenomenological behaviors but at the cost that they would limit the spatial and temporal grids. This extension is postponed to future work. Other 2D studies also include a disk in an idealized manner by introducing a resistive term which controls any excess in turbulence. Despite the decrease in the time step by the shorter interaction time scales in the disk (e.g., \citealt{casse2002magnetized,zanni2007mhd,2017MNRAS.468.3850S,Irelandetal2021,ireland2022effect}), the disk is treated more realistically. Computations on the largest domains were performed by \cite{anderson2005structure} with a jet followed up to 100 au and \cite{stepanovs2014-09modeling,stepanovs2014-11modeling} with a jet propagating to 140 au. These studies lack a stellar driven component, but include a resistive ingredient which is an important step toward more realistic disks. 

\subsubsection{Small scales}
For the inner part of the simulations, we reached a steady state in three stellar rotations only, equivalent to 42.6 days for the star of our simulations. This is not different from the results of \cite{sauty2022nonradial}, despite the more consistent treatment of the disk wind. The outer part, which contains the disk wind, takes longer to stabilize. Indeed, the initial perturbations due to the mixing of the spine jet and the disk wind need to relax. This takes time as the perturbations must exit from the outer boundary. In contrast, \cite{ZanniFerreira13} required over 50 stellar rotations, which amounts to more than 250 days, due to their faster rotating star. \cite{romanovaetal09,romanovaetal11} reached an equilibrium after 500 days, approximately 100 stellar rotations in their case. Consequently, our approach significantly shortens the relaxation timescale in the inner part even if the outer region has not yet stabilized. In more recent simulations by \cite{Irelandetal2021,ireland2022effect}, steady state was achieved after only 2 to 3 rotations of the central star, possibly because they included a high-velocity, low-density stellar jet component. It is worth noting that these simulations used smaller grids and/or lower resolutions. Another factor contributing to our shorter simulation time could be our initial conditions, which are closer to an equilibrium state, since we use analytical solutions for the inner spine jet and the outer disk wind.

\subsubsection{Large scales}
%\textcolor{magenta}{ (CS: I have a problem with this whole section:  where do you speak of your own work? It looks like a review of other people work. How does this apply to you own work?)}
According to BP82, cold disk models with $\kappa = 0.1$ should have $\lambda \sim 10 - 30$ for $\xi^{'}_0 = B_r/B_z \sim 1.2 - 1.5$ (see their Fig.2). We, on the other hand, obtain lever arms that depend on the heating used in the disk. For this reason, even though the magnetic profiles, mass loads and magnetization of the disk are imposed as boundary conditions, the heating modifies the outflow dynamics. It provides extra pressure to accelerate plasma above the disk. By doing so, the outflow does not need to have high lever arms as cold winds. It does not need such high lever arms to efficiently extract a significant part of the angular momentum.

The shock structures observed in our simulations seem to be a general by product of MHD collimation. They have been proposed as an outcome to self similar models of jet launching  for nonrelativistic jets \citep{gomez1993origin,ferreira1997magnetically} and for relativistic jets \citep{polko2010determining}.
%\textcolor{magenta}{I think there is an earlier reference on that by Gomez de Castro and Pelletier/pudritz}
The recollimation process is induced hoop-stress by the magnetic field. This is shown in \cite{ferreira1997magnetically} with self-similar cold models. It has been confirmed by 2D time-dependent simulations of \cite{hervet2017shocks} for the relativistic case and \cite{jannaud2023numerical} for nonrelativistic jets (shocks between $z \sim 70 - 250~\rm{au}$). 
%\textcolor{magenta}{CS a mention to Hervet et al 2017 for the relativistic case no?}
This force makes cold jets move toward the axis, creating standing recollimation shocks. This mechanism should work around different astrophysical objects, no matter the external environment. The authors of \cite{jannaud2023numerical} propose that such collimation shocks have not been seen before due to, either, lower physical scales or smaller times scales. Same as for analytical studies, they find that the recollimation position depends on the mass loading. the higher the mass loading, the closer the shocks are to the source. The key quantity that shapes the asymptotic of the jet is $B_z \propto r^{x-2}$. We also observe ins our simulations standing recollimation shocks closer to the central engine, around $z \sim 15 \ \rm{au}$ (see Sec.\ref{sec:standing recollimation}). Yet, there are three main differences in the present work. (i) First, the nature of the spine jet is different. \cite{jannaud2023numerical} minimize the contribution of the spine jet to isolate the collimation dynamic of the jet emitting disk. We, on the other hand, choose a spine jet solution that reproduces the key features of RY Tau micro-jet. (ii) Second, the disk wind of our simulations is not cold as we map the local heating rate. (iii) Third, we take into account the launching region and the disk produces a jet below $0.6 \ \rm{au}$.

However, despite the differences, the shocks that appear in \cite{Matsakosetal2008,Matsakosetal2009,stute2008stability,stute20143d,jannaud2023numerical} and in this work are all of the same nature. They are weak, oblique and with low compression factors between $\sim 1.5 - 2.5$, even if the launching at the base differs. In the case of \cite{Matsakosetal2008,Matsakosetal2009} the flow is already super fast at launching, and the magnetic field lines are being focused on the axis. For \cite{Matsakosetal2008,Matsakosetal2009} and \cite{stute2008stability}, the collimation is partly caused by boundary condition effects, and partly because the internal jet region is not strong enough to withstand the outer jet. \cite{jannaud2023numerical} assume a supersonic ouflow. 
%\textcolor{magenta}{CSnext sentence is unclear} 
Self-similar computation of the disk outflow find subsonic, but super-slow magnetosonic velocities \citep{ferreira1997magnetically}. All components in our simulations start in the sub slow magnetosonic region. 

Let us consider the parameters that we did not vary in this study, namely, $x$, $\mathcal{K}$ and $\mu$, which corresponds to the \cite{blandford1982hydromagnetic} (BP82) parameters. This helps the comparison with results presented in other publications. We modified the outflow conditions through heating, but other methods are possible as will be discussed below. The parameter $x=3/4$ is the same as what was used by BP82 for the magnetic field distribution. $\mathcal{K}$ is the mass-loading parameter and $\mu$, the ratio of thermal to magnetic pressures. The normalization is somewhat different from VTST00. In terms of BP82 formulation, we have $\kappa = 0.1$ and $\mu = 0.07$. These parameters are constant along the disk surface.

First, the parameter $x$, which controls the magnetic profile, controls the collimation degree of the jet. Naively, the more we increase $x$, the wider the jet. \cite{fendt2006collimation} performed 40 simulations to study the influence of the magnetic field profile on the collimation. They chose $\kappa \propto r^{3/2 - 2x}$, using the same boundary condition as \cite{ouyed1997numericala,ouyed1997numericalb}. $\mu$ was varied between science runs, as the inverse of $\kappa$. It has a increasing profile as $\varpi$ increases. They also explored the evolution of collimation given different density profiles. The grid they used to perform simulations is $[z,\varpi] = [300,150]$ in units of $10 R_\odot$ equivalent to $z \times \varpi = 13.3 \times 6.7 \ \rm{au}$. No spine component was considered in their work. \cite{fendt2006collimation} confirmed the role of the parameter $x$ on collimation. A decrease in the magnetic profile results in a decrease in collimation, irrespective of the density profile. Above $x>1.6$, they obtain no steady jet, and a wave shape is seen to propagate along the outflow. This observation is confirmed by more recent work, where for $x \geq 1$ time scales become too large to reach a steady state in a realistic computation time \citep{jannaud2023numerical}. In the same spirit, \cite{pudritz2006controlling} showed by varying $x=1, 3/4, 1/2, 1/4$, with a similar $\kappa$ profile as before, that a smaller $x$ leads to a less collimated jets. Although, they attribute the collimation property of the jet in their simulation to the current distribution, it is controlled by the mass load. CS check last sentence The expression of the azimuthal magnetic field is as follows, $B_\phi \propto r^{-1/2 - x}$. For cases with $x=1$ and $x=3/4$, the jet collimates, since $B_\phi$ is larger by decreasing $\kappa$. Conversely, the jet decollimates when $\kappa$ is increased. %\textcolor{magenta}{CS: check last sentences, not clear}

\cite{anderson2005structure} explore the jet evolution by changing the mass load from the disk for a cold jet model with $x=3/4$. They choose a constant $\kappa$ along the disk boundary. Its value goes from $6.3 \cdot 10^{-4}$ to 19, and outflow occurs in a distance of $1 \rm{au}$ in the inner part of the disk. In this region $B_z$ is reduced to minimize its role in shaping the poloidal field. 
%\textcolor{magenta}{CS: I do not understand the previous sentence}
They use a cylindrical grid $[z,\varpi] = [100,100] \ \rm{au}$ with $256 \times 256$ cells. They report a wider jet as $\kappa$ decreases. For a $\kappa$ larger than unity, they do not reach a steady state, and the magnetic field oscillates. This trend is in agreement with work done on steady jet theory (e.g., \cite{pelletier1992hydromagnetic,ferreira1997magnetically}). When a jet is heavily mass loaded, it affects its ability to produce a super Alfv\'enic outflow. Magnetic driven winds are possible up to $\kappa \sim 1$ producing a lever arm of $2$ \citep{blandford1982hydromagnetic}. Higher mass loadings bring the Alfv\'en surface closer to the disk, which in turn means the magnetic field has to have more inclination.

\subsubsection{Disk interior}
It is clear that platform simulations can only describe the accretion disk as a mere boundary condition for the wind. This in turn overlooks disk physics - not only pertaining to diffusivity and viscosity, but also the origin of the magnetic field. Simulations including the disk physics, \cite{zanni2007mhd}, find that increasing the resistivity reduces the ejection efficiency. The energy generated from the accretion is released in the jet and produces more powerful jets the less dissipation there is. In other words, the magnetic field applies a torque on the disk which will modify the efficiency of accretion. The evolution of the magnetic field leads to the readjustment of the jet itself. \cite{zanni2007mhd} characterize the effect of Ohmic heating on the ejection efficiency and find that an increase in the thermal energy inside the disk allows more mass loading in the jet. Our simulations are in agreement in that regard but for the outer part of the disk.

An important question to consider when studying an accretion disk is the origin of the magnetic field. Disk dynamos have been suggested to account for the large scale magnetic field having a direct effect on the evolution of the system. \cite{mattia2020magnetohydrodynamica} for instance consider the effect of a non-scalar dynamo tensor which was later extended to include analytical turbulent dynamo model with both magnetic diffusivity and a turbulent dynamo term \citep{mattia2020magnetohydrodynamicb}. They succeed at connecting both terms with the Coriolis number $\Omega^*$, related to the rotation frequency and the turbulence correlation time. As the Coriolis number increases so does the accretion rate. In other terms, accretion increases with the strength of the mean-field dynamo (refer to \citealt{mattia2020magnetohydrodynamicb} for more details). Even if we do not include diffusivity in our simulations, the magnetic field inversion in the nonstationary state depicted in Fig.\ref{fig:ME production} is similar to those present in their simulations.  When the Coriolis number is low  the disk is more turbulent and magnetic loops form. We argue in our case that without sufficient heating the diffusivity is not enough to amplify the magnetic field and have collimated ejections.
In \cite{mattia2022jets}, the same authors studied the various quenching of the dynamo and the diffusivity. In particular, they show that for high Coriolis numbers, the higher the magnetic field, the lower the acceleration and the higher the collimation. This is somehow similar to our simulations.  Decreasing the total heating enhances the influence of the magnetic field and the collimation while reducing the terminal jet speed.

\subsection{Comparison to observations}
Class II stars are studied using spectro-imaging, as well as forbidden emission lines and optical lines like $H_\alpha$ to trace the wind and jet behavior.  
From these lines, mass loss rates can be estimated. They, nonetheless, greatly depend on the geometry of the wind and its temperature \citep{fang2018new}. Through optical tracers the ratio of mass loss rates to accretion rates is $ \dot{M}_{wind}/\Dot{M}_{acc} \sim 0.1$ \citep{bouvier2002constraints}. This ratio stays constant throughout all evolutionary stages of young stars. This highlights the strong link between accretion and ejection \citep{konigl2000protostars}. In Tab. \ref{tab:simID}, we list the wind mass loss rates, the accretion rates and the ratio of mass loss to mass accretion rates of our simulations. All these quantities are calculated once a steady state is reached. As we increase the mass loss rate from the disk wind, the ratio of the stellar to the disk wind mass loss rates decreases. The stellar value remains nonetheless significant. 

Differentiating disk wind or stellar-driven material through observations is not possible for the jet itself, as it is not spatially resolved at the base. It follows that in the cases where a micro-jet is present, the separation between the stellar and the disk component is rather difficult. 
For RY Tau, for instance, a small jet confined to the axis is observed and attributed to the inner most part of the jet \citep{StongeBastien08}. Thus, even if the spine jet contribution is not known, it is expected to be important compared to the wind mass loss. In this study, the mass loss of the star is between $\sim 20 - 80 \%$ of the total mass loss. Overall, the micro-jet can be explained by the stellar component while the large scale jet is mainly comprised of disk wind. 

\cite{gomezetal01} used UV semi-forbidden lines as tracers to determine the physical properties of plasma near the base of the jet for RY Tau and RU Lup. They analyzed the high resolution spectra of $\left. C_{III}\right]_{1908}$ and $\left. Si_{III}\right]_{1892}$. They noted that the velocities of these lines are comparable to those of optical lines. The luminosity of the optical and semi-forbidden lines are also comparable. Looking at the emission of RY Tau in terms of semi-forbidden lines, the authors concluded they are not associated to the accretion shocks and should be produced farther than 2 stellar radii. Given the semi-forbidden lines are spatially broader than the optical lines, they argued that the observed profiles are due to shock structures close to the base of the jet. Provided the plasma is collisional and originating from the disk, they constrained the temperature of the emission region, $4.7<\log \ T_e(K)<5.0$, and the electronic density, $9<\log \ n_e(\rm{cm^{-3}})<11$. In a follow up paper, \cite{gomezetal07} further constrained the physical properties, given by the semi-forbidden lines, and determined that the outflows can not be of disk-origin only. Their argument is that the jet should be produced by the contributions of several outflows from atmospheric open-field structures like those observed in the Sun. The refined temperature for the shocks is $\log \ T_e(K) = 4.8$, and $9.5<\log \ n_e(\rm{cm^{-3}})<10.2$. The shocks we obtain in our simulations have temperatures of the same order as they presented. However they lack the density to be linked to UV emissions. The plasma density in the shocks of our simulations is two orders of magnitude lower than what is expected in the UV shocks. 

To have the necessary density the shock has to occur between $0.1$ and $0.45 \ \rm{au}$, which is consistent with the conclusion of \cite{gomezetal01}. The shock layer is inferred to occur closer than $\sim 38 R_\star$ ($\sim 0.5 \ \rm{au}$). Thus, the shocks in our simulations cannot be stricltly related to the UV shocks. However, they clearly show that jets with knots close to the star such as the ones we present in this article are of stellar origin. \cite{StongeBastien08} reported for the first time a jet in $H_\alpha$, which is also expected to be of stellar origin. The knots presence span from $\sim 200 \ \rm{au}$ to $4000 \ \rm{au}$. In our model the knots would be created at the shock layer and start migrating along the jet.

%"B. Tabone et. al. 2020 : Constraining MHD disk winds with ALMA Apparent rotation signatures and application to HH212" --> They discuss the lever arm observationally.

\subsection{Possible heating sources}
In the disk wind regions, the chemical state and thermal energy balance are influenced by stellar UV and X-ray flux. Accurately determining the gas temperature is crucial for calculating pressure gradients, wind flows and interpreting line emissions. Gas temperature is affected by heating and cooling processes, including collisions with dust grains, line emissions from ions, atoms, molecules and molecular abundances shaped by chemical reactions. However, including the microphysics of the heating rate is postponed to a later study. The uncertainty on the physical processes taking place in and above the disk are still not sufficiently well constrained to have a definitive answer.

At the disk surface where the wind originates, factors like Far UV photon-induced grain heating, conversion of binding energy during molecular hydrogen formation and excess electron energy from photo-dissociation and photo-ionization contribute to the gas heating. In addition, there is a  promising source which is the energetic particles from flares. They may serve as another important ionizing and heating source \citep{brunn2023ionization}.
Above the disk, magnetized jets and winds experience heating through nonideal MHD processes like Alfv\'en wave damping, Ohmic heating and ambipolar diffusion, especially at altitudes where photo-heating is less effective \citep[e.g.,][]{wang2019global}. This paper, focuses however on winds where the heating does not evolve with the fluid, underestimating the importance of the retro-action of heating sources on jet/wind generation. 

\section{Conclusions}

%\textcolor{magenta}{CS I think you can strenghten a bit the conclusions by making a more extended summary of your results}

We conducted 2.5D simulations of a jet showing key features from RY Tau, surrounded by a disk wind solution that reproduce outflows coming from a standard accretion disk. The jet is nonrelativistic, and the disk has a Keplerian rotation velocity profile. We mapped and parametrized the heating rate computed using the analytical solution quantities, namely, velocity and pressure. The outflows evolve, taking into account the heating map. To control the map, we added two parameters, with which we show a clear dependence of the jet shape at large distances with the heating rate at the base of the launching region. We find that the higher the heating rate, the less collimated the jet is. We distinguished between three cases using the measured opening angle of the spine at $10 \ \rm{au}$. While our definition of the jet radius remains ambiguous, it enables a consistent ordering between the different jet morphologies.

We showed, in the case of nonsteady jets, that a periodic behavior is possible. 
%\textcolor{magenta}{CS next sentence does not make sense, be more clear if you speak of the 2 states of RY Tau and refer to it as well} 
We showed that the jet can have two different states. The first may match the quiescent mode of RY Tau \citep{petrovetal19}, and the second one may match the other phase eruptive. These phases alternate in time. The eruptive phase occurs whenever the jet is perturbed by flares. During the flares, reconnection events occur. Matter is expelled due to the buildup of magnetic pressure. A stable period follows during which the length depends on the heating. Conversely, the quiescent phase takes place in between the mass ejection episodes. For the case we have shown, the simulation runs up to 14 years and we observed two events with this type of nature. Each occurrence is followed by a normal shock that moves outward along the axis. Observationally, the shocks are similar in nature to the  $H_\alpha$ optical lines observed by \cite{StongeBastien08}. Changing the heating map would change the period and strength of magnetospheric ejections, but the extent to which the period varies has not been studied. Even with a static heat map, we observed nonsteady cases with stable variations taking place over long periods (around $1$ to $7$ years)  

Out of the effects heating has, there are two power laws linking the jet-launching radius and the mass-loss rate to the heating rate. The radius from which the disk jet originates follows the power-law $\varpi_{JED} \propto Q^{-0.25}$, and the mass loss has the power-law $\dot{M}_{JED} \propto Q^{-0.7}$. 
Using these power laws, we further deduced a third link between the launching radius and the mass-loss rate, $\varpi \propto \dot{M}_{JED}^{0.33}$. This result is comparable to what was found by \cite{anderson2005structure} ($\varpi_j \propto \dot{M}_{\varpi_j}^{0.1}$). The main difference arises from the fact that \cite{anderson2005structure} used a cold wind with no spine jet being considered. Thus, the transverse force equilibrium is different.

Although unexpected, we obtain standing recollimation shocks below $100 \ \rm{au}$. These shocks may not be linked to the shocks observed in UV semi-forbidden lines \citep{gomezetal01,gomezetal07} because this would require them to occur closer to the source. Such a configuration may exist in a parameter space region we have not explored yet. However, the main takeaway from their study is that the shocks are of stellar origin, which is what we have observed in our simulations.
In any case, they match the shock structures found in \cite{StongeBastien08}, and previous numerical works, for example, \citep{Matsakosetal2009} and \cite{jannaud2023numerical}. The novelty resides in using a spine jet that matches the observations of the RY Tau micro-jet and extracts enough kinetic energy to maintain a constant rotation rate of the star. We also implemented a heating model that gives three jet shapes that can be compared to observations to identify the most probable heating values around a YSO.

\begin{acknowledgements}
We acknowledge financial support from “Programme National de Physique Stellaire” (PNPS) and from "Programme National des Hautes Energies" (PNHE) of CNRS/INSU, France. The authors would like to thank the anonymous referee for his fruitfull comments and suggestions. We also thank Andrea Mignone and the PLUTO team for the possibility to use their code. We thank Titos Matsakos for enlightening discussions over his model. The authors also acknowledge the support of France Grilles for providing computing resources on the French National Grid Infrastructure. CM would like to thank Gameiro Filipe and Lima João for the fruitful conversation and comments. CS thanks LUPM for hosting him during his long term visit (CRCT and délégations CNRS). We thank the support of IN2P3 via the INTERCOS project.
\\
Software: PLUTO (Mignone et al. 2007), Python 3.9.
\end{acknowledgements}

%\bibliography{bibliography}
%\bibliographystyle{aa}

\appendix
%\section{Simulation ID}

%\begin{table}[ht]
%    \centering
%    \caption{A list of the simulations performed with analytical radial boundary conditions. Except for R0\_I2 and R05\_I2, all simulations reach a steady state.}
%    \bgroup
%    \def\arraystretch{1.5}
%    \begin{tabular}{lcc}
%    \hline
%    \hline
%         Simulation ID & $R_c \ (AU)$ & $I_h$ \\ 
%         \hline
%         R0\_I1 & $0$ & $10^{-1}$ \\ 
%         R0\_I2 & $0$ & $10^{-2}$ \\ 
%         R05\_I1 & $0.5$ & $10^{-1}$ \\ 
%         R05\_I2 & $0.5$ & $10^{-2}$ \\ 
%         R1\_I1 & $1$ & $10^{-1}$ \\ 
%         R1\_I12 & $1$ & $0.5 \ 10^{-2}$ \\ 
%         R1\_I2 & $1$ & $10^{-2}$ \\ 
%         R1\_I22 & $1$ & $0.5 \ 10^{-2}$ \\ 
%         R1\_I23 & $1$ & $0.3 \ 10^{-2}$ \\  
%         R2\_I13 & $2$ & $0.3 \ 10^{-1}$ \\ 
%         R2\_I2 & $2$ & $10^{-2}$ \\ 
%         R2\_I3 & $2$ & $10^{-3}$ \\
%         R2\_I4 & $2$ & $10^{-4}$ \\
%         R3\_I2 & $3$ & $10^{-2}$ \\ 
%         R3\_I3 & $3$ & $10^{-3}$ \\ 
%         R3\_I4 & $3$ & $10^{-4}$ \\
%         R4\_I3 & $4$ & $10^{-3}$ \\
%         R4\_I4 & $4$ & $10^{-4}$ \\ 
%         \hline
%    \end{tabular}
%    \egroup
%    \label{tab:simID}
%\end{table}

\section{A map of jet and wind tracers}\label{tarcers}
\begin{figure}[htbp]
    \centering
    \includegraphics[width=.5\textwidth ]{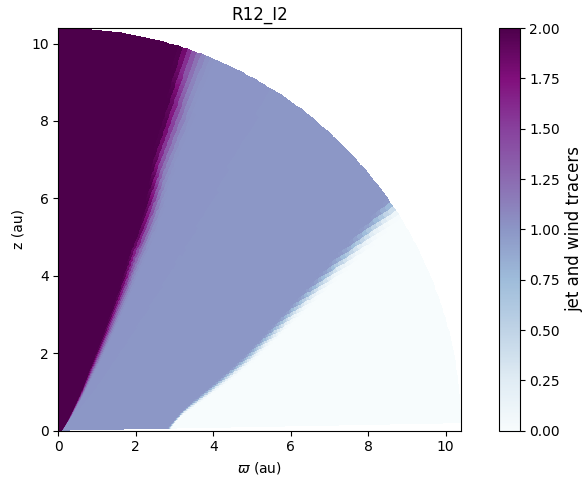}
    \caption{2D map of jet and wind tracers. The dark purple indicates the evolution of the spine and the light purple shows the disk wind coming from below $3 \rm{au}$ of the disk surface. We plot the final time.}
    \label{fig:tracers}
\end{figure}

From the map shown in Fig.\ref{fig:tracers}, we deduce the opening angle of the dark purple region. In cases where the spine varies slightly during the evolution, we do a time-average of the measured opening angles.    

\section{Additional table}
To gauge how simulations evolve with different disk densities, we tried three cases as presented in Tab.\ref{tab:den norm}. More details about the results are given in Sec.\ref{var rho}
\begin{table*}[htbp]
\centering
\caption{Table of the mass-loss and wind to accretion ratio for different density normalization factors at $t=0$. The mass loss of the star in all cases is $3.0 \cdot 10^{-9}$ $\Msunyr$. The ratio of stellar mass loss rate to accretion rate is, also, common to the three cases, $\lvert \dot{M}_{stel}/\Dot{M}_{acc} \rvert \simeq 0.23$.}
\bgroup
\def\arraystretch{1.35}
\begin{tabular}{c c c c c c c c c}
 \hline
 \hline
 $\text{name}$ & $l_\rho$ &  $l_B$ & $d_m$ & $\alpha_m$ & $\dot{M}_{wind} (10^{-9}\Msunyr)$ & $\dot{M}_{acc} (10^{-8}\Msunyr)$ & $ \lvert \dot{M}_{wind}/\dot{M}_{acc} \rvert$ & $\dot{M}_{stel} / \dot{M}_{wind}$\\
 \hline
 R3A1 & 0.004 & 0.013 & 2 & 1 & $6.9$ & -1.3 & 0.53 & 0.43\\
 R3A2 & 0.014 & 0.025 & 2 & 1 & $3.0$ & -1.3 & 0.23 & 1\\
 R3A3 & 0.044 & 0.044 & 2 & 1 & $1.8$ & -1.3 & 0.14 & 1.7\\
 \hline
\end{tabular}
\egroup
\label{tab:den norm}
\end{table*}

\section{ADO-ASO models}\label{annex param}
Assuming axi-symmetry, steady state, and self-similarity, the ideal magnetohydrodynamics (MHD) equations reduce to a system of coupled ordinary differential equations (ODEs) in spherical coordinates. The equations are then  solved by providing the values of key functions for each analytical model.

%Axi-symmetry, steady state and self-similarity assumptions reduce the complexity of the ideal MHD equations to a set of coupled ODEs in spherical coordinates. The equations are solved numerically by providing the values of key functions for each analytical model.

For the ADO solution, physical quantities are provided by key functions $G_d(\theta)$, $M_d(\theta)$and $\Psi_d(\theta)$,
\begin{equation}\label{norm}
\frac{\rho_{D}}{\rho_{D_\star}}= \alpha_D^{x-3/2}\frac{1}{M^2},
\end{equation}

\begin{equation}
\frac{P_{D}}{P_{D_\star}}= \alpha_D^{x-2} \frac{1}{M_{D}^{2\gamma}},
\end{equation}

\begin{equation}\label{norm1}
\frac{\boldsymbol{B}_{D_p}}{B_{D_\star}} = 
- \alpha_D^{\frac{x}{2}-1} \frac{1}{G_{D}^2}
\frac{\sin \theta}{\cos \left(\psi + \theta \right) }
\left(\sin \left(\psi+\theta \right)\hat{\boldsymbol{r}} + \cos \left(\psi+\theta \right) \hat{\boldsymbol{\theta}}\right), 
\end{equation}

\begin{equation}\label{norm2}
\frac{\boldsymbol{V}_{D_p}}{V_{D_\star}}= 
- \alpha_D^{-1/4} \frac{M_{D}^2}{G_{D}^2} \frac{\sin \theta}{\cos \left(\psi +\theta \right) }
\left(\sin \left(\psi+\theta \right)\hat{\boldsymbol{r}} + \cos \left(\psi+\theta \right) \hat{\boldsymbol{\theta}}\right),
\end{equation}

\begin{equation}\label{norm3}
\frac{B_{D_\varphi}}{B_{D_\star}}=-\lambda_{D}  {\alpha_D}^{\frac{x}{2}-1} \frac{1-G_{D}^2}{G_{D}\left(1-M_{D}^2 \right)},
\end{equation}

\begin{equation}
\frac{V_{D_\varphi}}{V_{D_\star}}=\lambda_{D}  {\alpha_D}^{-1/4} \frac{G_{D}^2-M_{D}^2}{G_{D} \left(1-M_{D}^2 \right)},
\end{equation}
$M_D^2(\theta)$ is defined as the square of the poloidal Alfvén Mach number and is given by,
\begin{equation}
M_D^2(\theta) = 4\pi \rho \frac{\boldsymbol{V}^2_{D_p}}{\boldsymbol{B}^2_{D_p}}\;, 
\end{equation}
The next key function is $G_D(\theta)$, the cylindrical cross section of a flux tube defined by the flux function $\alpha_D$. Finally, $\Psi(\theta)$ is the angle between a particular poloidal field line and the radial cylindrical axis.
The nondimensional magnetic flux function $\alpha_D$ is given by,
\begin{equation}
    \alpha_D = \frac{\varpi^2}{\varpi^2_{\star} G^2_{D}}
\end{equation}

The ASO solution is, on the other hand, described with the help of the following functions, $G_s(R)$, $M_s(R)$, $F_s(R)$and $\Pi_s(R)$,
\begin{equation}\label{pressure}
P_{S}(R,\alpha) = {\frac{1}{2}} \rho_* V^2_* \Pi[1+ \kappa_{S} \alpha]
+P_{naught}
\end{equation}
\begin{equation}
\rho_S(R,\alpha) = {\frac{\rho_*}{M_{S}^2}} (1 + \delta \alpha)
\end{equation}
\begin{equation}
\label{Br}
\frac{\boldsymbol{B}_S{_p}}{B_{S_*}} = {\frac{1}{G_{S}^2}} \bigl(\cos{\theta} \ \hat{\boldsymbol{r}} - {\frac{F}{2}}\sin\theta \ \hat{\boldsymbol{\theta}} \bigr)
\end{equation}
\begin{equation}
\label{Vr}
\frac{\boldsymbol{V}_S{_p}}{V_{S_*}} = {\frac{M_{S}^2}{G_{S}^2}} { \frac{1}
{\sqrt{1+\delta \alpha(R,\theta)}}} \bigl(\cos{\theta} \ \hat{\boldsymbol{r}} - {\frac{F}{2}}\sin\theta \ \hat{\boldsymbol{\theta}} \bigr)
\end{equation}
\begin{equation}
\label{Vphi}
{\frac{V_S{_\varphi}}{V_{S_*}}} =   {\frac{\lambda_{S}}{G_{S}^2}}
{ \frac{G_{S}^2 - M_{S}^2}{1- M_{S}^2}}
{\frac{R\sin\theta}{\sqrt{1+ \delta \alpha(R,\theta) }} }
\end{equation}
\begin{equation}
\frac{B_S{_\varphi}}{B_{S_*}} = -  {\frac{\lambda_{S}}{G_{S}^2}} {\frac{1 - G_{S}^2}{1 - M_{S}^2 }} {R\sin\theta}
\end{equation}
This model also uses certain key functions to provide a solution for the stellar jet component. These functions are $M_S(R)$, $G_S(R)$, $F(R)$, and $\Pi(R)$. Similarly to the ADO model $M_S(R)$ and $G_S(R)$ have the same definitions, but depend on R. $\Pi(R)$ is the dimensionless pressure. $F(R)$ is analogous to $\Psi_D(\theta)$ as it represents a logarithmic expansion factor, which measures the angle of a field line projected on the poloidal plane along the radial direction. It is written as,
\begin{equation}\label{F(r)}
    F(R) = \frac{\partial \ \ln\alpha_S}{\partial \ \ln R} = 2\left( 1 - \frac{\partial \ \ln G_S}{\partial \ \ln R} \right)\;.
\end{equation}

The nondimensional magnetic flux function $\alpha_S$ is given by,
\begin{equation}
    \alpha_S = \frac{\varpi^2}{r^2_{\star} G^2_{S}}
\end{equation}

The parameters used can be linked to the solutions as follows,
\begin{equation}
    \mu = \frac{2P_{D}}{B_{D}},\quad \lambda_s = \frac{G^2_{a}}{G^2(R_{ss})} ,\quad \lambda_d = \frac{G^2_{a}}{G^2(R_{ds})},
\end{equation}
on the Alfvén surface the key function $G$ is equal to zero. The subscripts a, ss, and ds are for Alfvén surface, stellar surface, and disk surface, respectively.
%-------------------------------------------------------------------

\end{document}